\documentclass[bibyear]{aa}  
\usepackage{graphicx}
\usepackage{txfonts}
\usepackage{multirow}
\usepackage{natbib}
\bibpunct{(}{)}{;}{a}{}{,} % to follow the A&A style
\usepackage[]{hyperref}
\def\xmm{{XMM-{\it Newton\/}}}
\def\chandra{{{\it Chandra\/}}}
\def\nus{{{\it NuSTAR\/}}}

\begin{document}

   \title{The role of molecular gas in the nuclear regions of\\
   IRAS 00183-7111}

   \subtitle{ALMA and X-ray investigations of an ultraluminous infrared galaxy}

   \author{I. Ruffa\inst{1,2}
          \and
          C. Vignali\inst{1,3}%\fnmsep\thanks{Just to show the usage
          %of the elements in the author field}
          \and
          A. Mignano\inst{2}
          \and
          R. Paladino\inst{2}
          \and
          K. Iwasawa\inst{4,5}
          }

   \institute{Dipartimento di Fisica e Astronomia, Alma Mater Studiorum, Universit\`{a} degli Studi di Bologna, via Gobetti 93/2, 40129 Bologna, Italy\\
             \email{i.ruffa@ira.inaf.it}
             \and
             INAF - Istituto di Radioastronomia, via P. Gobetti 101, 40129 Bologna, Italy
             \and
             INAF - Osservatorio di Astrofisica e Scienza dello Spazio di Bologna, via Gobetti 93/3, 40129 Bologna, Italy
             \and
             Institut de Ci\`{e}ncies del Cosmos (ICCUB), Universitat de Barcelona (IEEC-UB), Mart\'{i} i Franqu\`{e}s, 1, 08028 Barcelona, Spain
             \and
             ICREA, Pg. Llu\'{i}s Companys 23, 08010 Barcelona, Spain
             %\thanks{The university of heaven temporarily does not
                     %accept e-mails}
             }

   \date{Received accepted}

% \abstract{}{}{}{}{} 
% 5 {} token are mandatory
 
  \abstract
   {%IRAS 00183-7111 (z=0.329) is one of the most luminous Ultra Luminous Infrared Galaxy (ULIRG) detected by IRAS ($L_{bol}=9\times10^{12}~L_{\odot}$). Multi-wavelengths observations reveal the peculiarity of this source: it seems to have been caught in the transition period between a post-merger powerful starburst and the birth of a quasar. A significant ULIRG fraction seems to host a heavily obscured AGN in their nuclei, and this fraction increases at increasing infrared luminosity. The detection of the most obscured objects is crucial to shed light on the obscured accretion phase in black hole growth, the AGN/host-galaxies co-evolution issue, and eventually estimate the contribution of these sources to the cosmic X-ray background.
   }
   {We present a multi-frequency study of the ultraluminous infrared galaxy (ULIRG) IRAS 00183-7111 ($z=0.327$), selected from the Spoon diagnostic diagram as a highly obscured active galactic nucleus (AGN) candidate. ALMA millimetre and X-ray observations are used;  the main aim is to verify at what level the molecular gas, traced by the CO, may be responsible for the obscuration observed at X-ray energies. Theory and observations both suggest that galaxy-scale absorption may play a role in the AGN obscuration at intermediate (i.e. Compton-thin) column densities.}
   { We calibrated and analysed ALMA archival Cycle 0 data in two bands (Bands 3 and 6). The X-ray properties of IRAS 00183-7111 were studied by reducing and analysing separately archival \chandra\ and \xmm\ data; recently acquired \nus\ spectra were first examined individually and then added to the \chandra\ and XMM spectra for the broad-band (0.5$-$24 keV, observed frame) analysis.}
   {We derived a molecular gas column density of $(8.0\pm0.9)\times10^{21}$~cm$^{-2}$ from the  ALMA CO$_{(1-0)}$ detection, while the best-fit column density of cold gas obtained from X-ray spectral fitting is $6.8^{+2.1}_{-1.5}\times10^{22}$~cm$^{-2}$. The two quantities suggest that the molecular gas may contribute only  a fraction of the AGN obscuration; however, the link between them is not straightforward. The nuclear regions of IRAS 00183-7111 are likely stratified into different layers of matter: one inner and highly ionized by the strong radiation field of the AGN (as inferred from the high-ionization iron line found in the X-ray spectra), and one outer and colder, extending more than 5~kpc from the nucleus (as traced by the molecular gas observed with ALMA). The molecular gas regions also give rise to a vigorous starburst with SFR$\sim260\pm28$~M$_{\odot}$~yr$^{-1}$. The complexity of this nuclear environment makes it difficult to identify the origin of the AGN obscuration given the quality of the data currently available. Higher resolution observations in the millimetre regime are needed to deeply investigate this issue.}
   {}


   \maketitle

\section{Introduction}
Ultraluminous infrared galaxies (ULIRGs) are a class of galaxies defined in terms of their huge infrared luminosity, $L_{8-1000 \mu m}>10^{12}~L_{\odot}$ \citep[e.g. ][]{Sanders96}. ULIRGs were discovered by the Infrared Astronomical Satellite (IRAS) in 1983 and since then many efforts have been made to understand their origin and their overall characteristics. It is now well established that the source of radiation field inside ULIRGs is generally a combination of extreme starburst activity and highly obscured accretion on a pre-existing supermassive black hole \citep[e.g.][]{Genzel98}. However, the distinction between the two processes is difficult due to the high dust extinction of their nuclei, which leads to a substantial obscuration even in mid-infrared and X-ray regimes. Despite this, several spectroscopic signatures of starburst and/or AGN activity can be identified, thus in recent years, many efforts have been directed to disentangling the AGN and starburst components in ULIRGs \citep[e.g.][]{Veilleux02, Franceschini03, Farrah06, Imanishi10}. Most of these works represent systematic studies of large nearby samples of ULIRGs, observed in radio, mid-infrared, and X-ray regimes, and their conclusions seem to converge on some fundamental points:
\begin{itemize}
\item The majority of ULIRGs have optical spectra characterized by typical starburst features, and the fraction of ULIRGs with either type-I (unobscured) or type-II (obscured)   AGN spectra increases at increasing IR luminosity; however, most  ULIRGs with dominant AGN spectral features also show evidence of ongoing or recent star formation;
\item The AGN in ULIRGs can be classified as faint X-ray sources, with hard X-ray luminosity being usually $L_{2-10keV}\leq10^{42}-10^{43}$~erg~s$^{-1}$ \citep [e.g.][]{Franceschini03, Lonsdale06};  furthermore, ULIRG X-ray spectra typically show an excess of emission below 2~keV, which  in most cases can be  associated with thermal emission with kT$\sim$0.7~keV, related to  nuclear or circumnuclear starburst activity;
\item From a combination of mid-IR and X-ray spectral analyses, it was found that the environment surrounding the AGN component in ULIRGs is much richer in gas and dust than ordinary active galaxies, and the degree of absorption can be directly related to the starburst intensity, indicating a strong link between the two processes. 
\end{itemize}
The implications of this kind of study are fundamental to understanding the nature of local ULIRGs, and also to placing constraints on the interplay and mutual feedback between star formation and black hole (BH) accretion, which are the basic ingredients of galaxy formation and evolution \citep[e.g.][]{Zheng13}.\\
The most popular theoretical picture suggests that ULIRGs are triggered by major merger events between gas-rich galaxies which channel very large masses of gas and dust into the few hundred parsecs of their central regions \citep[e.g.][]{Sanders96, Veilleux02}. For this reason, ULIRGs are placed in an evolutionary sequence where the major galaxy mergers first result in a massive cool starburst-dominated ULIRG which is then followed by a warm\footnote{The definitions of `cool' and `warm'  are based on infrared colours:  a ULIRG is defined warm when $f_{25}/f_{60}>0.2$, where $f_{25}$ and $f_{60}$ are 25 and 60 $\mu$m IRAS flux densities, respectively.} ULIRG, as the QSO turns on inside and heats the surrounding layers. Finally, the QSO emerges in an optically bright phase, and it blows away the surrounding material, quenching both star formation and AGN activity. The resulting stellar system evolves as a passive spheroid \citep[e.g.][]{Lipari03,Hopkins08}. According to these models, a significant fraction of heavily obscured AGN are harboured in ULIRGs at late merger stages; for this reason, adopting a multi-wavelength approach and using the best data currently available is the best strategy for shedding light on the open AGN/ULIRG issue. This picture provides supporting evidence to the co-evolution framework;  what is still not clear is the  connection with the Unified Model scenario \citep{Antonucci93,Urry95}, which states that the observed differences within the AGN population have to be ascribed to different  orientations with respect to the line of sight. In this framework, type-I (unobscured) AGN  are nearly face-on objects, and the nucleus can be directly seen; type-II (obscured) AGN have larger viewing angles, thus the line of sight intercepts an asymmetric dusty structure known as the torus, whose inner radius (i.e. minimal distance from the BH) is a few parsec \citep[e.g.][]{Jaffe04}. Then, according to the Unified Model, a ULIRG hosting an optically bright AGN  viewed not completely face-on would appear as a type-II AGN.\\
However, over the years, the Unified Model scheme has been  discussed, tested, and revisited several times \citep[e.g.][]{Matt00,Elvis12,Bianchi12,Netzer15,Mateos16,Ramos17}. One intriguing issue concerns the origin and the properties of the obscuring medium in type-II AGN and its connection with the host--galaxy environment. Currently, there are some indications that suggest a more complex situation than that described by the Unified Model for type-II AGN: the suppression of  nuclear radiation may occur on a wide range of different spatial scales, varying on a galaxy-by-galaxy basis and making it difficult to  fit this complex scenario into a simple unification scheme \citep[e.g.][]{Netzer15}. In particular, it is now widely believed that the scales on which obscuration occurs may include at least three components: (1) an inner region, located at sub-parsec scales, usually associated with broad emission line clouds \citep[called the broad-line regions, e.g.][]{Risaliti09}; (2) an intermediate dusty obscurer, located on parsec scales, that reprocesses much of the AGN luminosity into thermal dust emission and is usually identified as the torus predicted by the Unified Model \citep[e.g.][]{Jaffe04}; (3) outer regions, up to kiloparsec scales, related to the host galaxy gas reservoir and/or dust lanes \citep[e.g.][]{Guainazzi05}.\\
The role of an obscuring medium on the host-galaxy scales is thought to be important at intermediate column densities (Compton-thin, i.e. absorbing column densities in the range $\sim10^{21}-10^{24}$ cm$^{-2}$), which in some cases are consistent with the optical reddening associated with the host-galaxy medium \citep[e.g.][]{Bianchi12}. In this regard, \citet{Matt00} proposed that the most obscured (Compton-thick, i.e. absorbing column densities exceeding $10^{24}$ cm$^{-2}$) objects are observed through the torus, while the Compton-thin objects are observed through a dust lane belonging to the host galaxy. Important differences can  especially be found for those sources showing clear signs of star formation activity, interactions, or merger events that channel large reservoirs of gas and dust into the AGN surroundings \citep[e.g.][]{Weaver01,Juneau13}. Therefore, a connection with the host galaxy environment is often invoked to explain the AGN obscuration, but its origin is far from being properly understood.\\
In this context, sources like ULIRGs represent unique laboratories to investigate the connection of the host-galaxy in the AGN obscuration: at the intermediate or late stages of their evolution both AGN and starburst components often contribute to the total power, and large amounts of gas and dust are channelled to the inner few hundred parsec. Our study is inserted in this framework.\\
In this paper we present a study of the multi-frequency properties of  the ULIRG IRAS 00183-7111, connecting ALMA millimetre observations with those at high energies. One of the main goals of our work is to verify whether  the gas (traced by the CO) can be considered responsible for the obscuration inferred from X-ray observations, and if so at what
level.
Similar studies could have implications not only in the framework of the co-evolutionary scenario, but also to test the scales on which AGN obscuration occurs, then verifying if the presence of a dusty torus surrounding the accreting black hole is not the only way to explain the line-of-sight obscuration in type-II AGN.\\
The paper is organized as follows: in Section~\ref{sec:Source selection} we describe the source selection criterion and the target multi-frequency properties; in Sections~\ref{sec:ALMA obs} and \ref{sec:ALMA results} we illustrate the ALMA observation properties, the main steps of data reduction, and the obtained results; in Section~\ref{X-ray obs} we describe the X-ray observations and spectral fitting; in Section~\ref{sec:discussion} we discuss our results before concluding in Section~\ref{sec:concl}.\\
Throughout this paper distances and luminosity have been computed assuming a $\Lambda$CDM cosmology: $H_{0}=70$~km~s$^{-1}$ Mpc$^{-1}$, $\Omega_{M}=0.27$, and $\Omega_{\Lambda}=0.73$. 
   
\section{Source selection: the case of IRAS 00183-7111}\label{sec:Source selection}
The 5$-$37~$\mu$m spectral properties of a sample of $\sim$100 ULIRGs, along with a sample of AGN, starbursts, and normal galaxies were analysed by \citet{Spoon07} using   observations  carried out by the Spitzer Space Telescope and the Infrared Space Observatory (ISO). In particular, \citet{Spoon07} introduced the comparison between the strength of the 9.7~$\mu$m silicate absorption feature and the 6.2~$\mu$m polycyclic aromatic hydrocarbons (PAH) emission feature as a powerful tool that can be used to investigate  the AGN/starburst component in ULIRGs, and plotted the two quantities into  the Spoon diagnostic diagram. The galaxies of the sample are divided into nine classes on the basis of their mid-infrared spectral properties. We selected IRAS 00183-7111 (hereafter I00183) from the 3A region of the Spoon diagram, where the absorption-dominated candidate sources are located; our purpose was to select a target in which we can maximize the possibility of finding a highly obscured AGN with limited  star formation activity.\\
I00183 is at z=0.327 and is one of the most luminous ULIRGs discovered by IRAS, with a bolometric luminosity of $9\times10^{12}$~L$_{\odot}$ \citep{Spoon09}, mostly radiated at far-infrared (FIR) wavelengths. I00183 has been optically classified as a type-II Seyfert galaxy by \citet{Armus89}; mid-infrared imaging in the K band by \citet{Rigopoulou99} revealed a disturbed morphology and a single nucleus; and long-slit spectroscopy \citep{Drake04} shows bright, highly disturbed, ionized gas extending 50~kpc east and 10~kpc west of the nucleus. Recently, \citet{Norris13} presented a R$+$I band image of I00183 obtained with the Australian Astronomical Telescope (AAT) in June 2012 which shows what appears to be either a  tidal tail or remnants from a recent merger. However, they argued that the high level of extinction that affects the nucleus ($A_{V}\geq90$) indicates that perhaps we are seeing the outer shells of the galaxy rather than the nucleus \citep{Norris13}.\\
I00183 is well known to host nuclear outflows of gas. A first fast outflow signature was observed by \citet{Heckman90} identifying a blueshifted [O{\sc iii}]$\lambda5007$ emission extending about 10~arcsec to the east of the nucleus. Nuclear outflows traced by the 12.81~$\mu$m [Ne{\sc ii}] and 15.51~$\mu$m [Ne{\sc iii}] lines were then observed by \citet{Spoon09}, with  clearly asymmetric and strongly blueshifted line profiles (FWHM$\sim3000$~km~s$^{-1}$), likely originating in a region $<3"$ east  of the nucleus. Line ratios are consistent with an origin in fast shocks ($v>500$~km~s$^{-1}$) in an environment with gas densities $n>10$~cm$^{-3}$, leading to consider that they may trace the initial stages of the disruption of the obscuring medium around a buried AGN \citep{Spoon09}. More recently, \citet{Calderon16} detected the OH doublet at 119$\mu$m purely in absorption, showing a significant blueshifted wing that may trace a molecular outflow with $v_{max}\sim1500$~km~s$^{-1}$.\\
A detailed analysis of the 4$-$27~$\mu$m spectrum of I00183 was presented and discussed by \citet{Spoon04}. The spectrum is dominated by broad absorption features, but  emission features have also been detected, such as the 11.2~$\mu$m PAH and the 12.7--12.8~$\mu$m blend of PAH and [Ne{\sc ii}] emission lines. In particular, from the strength of the 11.2~$\mu$m PAH feature, they evaluated that the starburst activity has to contribute to at least 30\% to the bolometric luminosity of the source. On the basis of the obtained results, \citet{Spoon04} suggested that the central dominant power source seems to be deeply buried behind two obscuring shells: the inner shell (on scales of less than 0.03 pc) is composed of a warm and dense gas (T$\sim720$~K, $n>3\times10^{6}$~cm$^{-3}$) that gives rise to CO absorption bands; the outer and colder shell is responsible for the absorption features of ice and silicates, and its column density was evaluated to be at least $10^{23}$~cm$^{-2}$. Observationally, the I00183 IR spectrum lacks the 7.65~$\mu$m [Ne{\sc iv}], 24.3/14.3~$\mu$m [Ne{\sc v}], and the [O{\sc iv}]  lines, which are distinctive features in AGN spectra \citep{Spoon09}, but overall  it has been  classified as a peculiar example of an extremely obscured AGN. The differences from a classical AGN template spectrum at optical and infrared wavelengths are attributed to the dense dust layers surrounding it. The presence of cold and dense molecular gas in the central regions of the galaxy was also indicated by weak ($\sim3\sigma$) ALMA detections of HCN, HCO$^{+}$, and HNC emission lines at $\sim1$~mm \citep{Imanishi14}; the detections were too weak to draw strong conclusions, but it is important to note that enhanced HCN emission is typical in AGN-dominated ULIRGs \citep{Imanishi09}.\\
Radio observations are generally unaffected by the heavy dust extinction that blocks the view at shorter wavelengths, thus allow us to see the
nucleus directly. The I00183 radio luminosity of $\sim10^{25.4}$~W~Hz$^{-1}$ at 4.8~GHz indicates that it is significantly more radio-luminous than expected from its  FIR emission; such radio-excess places it within the regime of high-luminosity (FRII-class) radio galaxies \citep{Roy97}. Very Long Baseline Interferometry (VLBI) observations of I00183 presented by \citet{Norris12} reveal the presence of a compact core-jet AGN with a double-lobed morphology and compact jets only 1.7~kpc long.\\
More recently, \citet{Mao14} presented an ATCA CO$_{(1-0)}$ detection (7$\sigma$) from which they derived a CO luminosity of 1.2$\times10^{10}$~K~km~s$^{-1}$~pc$^{2}$. This luminosity was then converted into a molecular gas mass of $1.0\times10^{10}$~M$_{\odot}$, assuming a H$_{2}$ mass-to-CO luminosity conversion factor, $\alpha$, standard for ULIRGs \citep[0.8; e.g.][]{Downes98}. Then, using the empirical CO-infrared luminosity relation \citep[e.g.][]{Carilli13}, they estimated a star formation rate (SFR) of $\sim220$~M$_{\odot}$~yr$^{-1}$, concluding that star formation accounts for only 14\% of the total power of this source.\\
The presence of a powerful AGN source in I00183 was definitely confirmed by \citet{Nandra07} through the detection of a 6.7~keV FeK line (Fe XXV) with a large equivalent width ($\sim1$~keV), indicative of reflected light from a heavily obscured object. \citet{Nandra07} also provided  an estimate of the star formation in I00183. They pointed out that the SFR derived from hard X-ray luminosity using the most conservative relation in \citet{Grimm03} would be >1.2$\times10^{4}$~M$_{\odot}$~yr$^{-1}$, while that converted directly from the total infrared luminosity \citep{Kennicutt98} would be about 10$^{3}$~M$_{\odot}$~yr$^{-1}$; this suggests that the source is  too luminous in hard X-rays to be solely powered by a starburst. On the other hand, using the method described in \citet{Ranalli03}, they derived a SFR of 310~M$_{\odot}$~yr$^{-1}$ from soft X-ray luminosity. Therefore, taking into account the results obtained by \citet{Nandra07} and \citet{Mao14}, the star formation contributes  $<$30\% to the bolometric luminosity of this source, in agreement with the value given in  \citet{Spoon04}.\\
All the observed characteristics seem to converge on the idea that I00183 hosts a powerful AGN in its nuclear regions and may have been caught in a transition period coincident with the initial stage of the formation of an optically bright quasar that has just started boring its way through the surrounding layers of gas and dust.

\section{ALMA observations}\label{sec:ALMA obs}
I00183 observations were taken during ALMA Early Science Cycle 0, between November 2011 and December 2012 in two frequency bands: Band 3~($\sim87$ GHz, project code: 2011.0.00034.S) and Band 6 ($\sim$270~GHz, project code: 2011.0.00020.S).\\
Band 3 data consist of three observations taken between 18 and 31 December 2012, for a total time-on-source of 78.1~min. The spectral configuration is composed of four different 1.875~GHz-wide spectral windows, divided into 128 channels of 15.6~MHz in width ($\sim45$~km/s); 87~GHz is the representative frequency. About 25 antennas of 12~m were arranged in a compact configuration, with baselines lengths ranging from 15 to 400~m.
The QSO J2225-049 was used as the bandpass calibrator, while the phase calibrator is J2157-694; either Neptune or Callisto was used for the absolute flux calibration.\\
Band 6 data were  taken in eight observing blocks between 26 November 2011 and 23 January 2012, for a total time on source of 215.14~min. The observations were  made in two different spectral configurations: the first  consists of two 1.875~GHz  spectral windows composed of 3840~$\times$~0.98~MHz ($\sim1$~km/s) channels, centred on a  representative frequency of 263~GHz; the second  consists of four 1.875~GHz spectral windows, with 3840~$\times$~0.98~MHz channels, and 270~GHz as the representative frequency. 18~$\times$~12m antennas were arranged in a compact configuration, with baselines ranging from 18 to 269~m.
Neptune was used for the absolute flux calibration of each execution block (except one, for which Callisto was used), while 3C454.3 and J2157-694 were used as bandpass and phase calibrators, respectively. Table~\ref{tab:ALMA observations summary} summarizes the ALMA observations properties.\\
The data reduction was conducted using CASA, version 3.4.2, performing a standard calibration procedure. Data re-processing is strongly suggested for Cycle 0 data, mainly because new flux model libraries were released after the archival data reduction, providing a more reliable flux calibration\footnote{\tiny{https://help.almascience.org/index.php?/Knowledgebase/Article/View/161/0/}}. We calibrated  each dataset separately, starting from the ASDM data format; once a reliable calibration was obtained, we combined data with the same spectral configuration into one calibrated dataset. Significant improvements were obtained by our data reduction with respect to that available from the archive; in particular, we can underline a more reliable flux calibration, thanks to the use of the latest model library releases. The images were finally produced using the CLEAN algorithm \citep{Hogbom74, Clark80, Schwab84}.
\subsection{Continuum imaging}
Continuum maps were produced using the CLEAN task in \textit{Multi Frequency Synthesis (MFS)} mode \citep{Rau11}, and both natural and Briggs weighting (robust=0.5); the first is used to maximize the sensitivity, the second to improve the spatial resolution maintaining a good signal-to-noise ratio (S/N). The image properties detailed below were measured in natural weighted maps; the quoted errors account for the $\sim$10\% flux calibration uncertainty of the ALMA data.\\
The 3.2~mm (87~GHz, Band 3) continuum is characterized by a synthesized beam FWHM of 3.01"$\times$2.05". Multiple rounds of phase-only self calibration were performed, yielding a 1$\sigma$ root-mean square (rms) noise level of 16~$\mu$Jy~beam$^{-1}$, with a peak flux density of 3.2$\pm$0.3~mJy~beam$^{-1}$ in the final map.\\
The 270~GHz (1.1~mm, Band 6) continuum image has synthesized beam FWHM of 2.35"$\times$1.19". The measured rms noise level is 15~$\mu$Jy~beam$^{-1}$, with a peak flux density of 1.7$\pm$0.2~mJy~beam$^{-1}$. At 263~GHz (Band 6) the continuum image has a synthesized beam FWHM of 2.48"$\times$1.31". The measured 1$\sigma$ rms noise level is 48~$\mu$Jy~beam$^{-1}$, with a peak flux density of 2.3$\pm$0.2~mJy~beam$^{-1}$.
\subsection{Line imaging}
\begin{figure*}[htbp]
\centering
\includegraphics[scale=0.5]{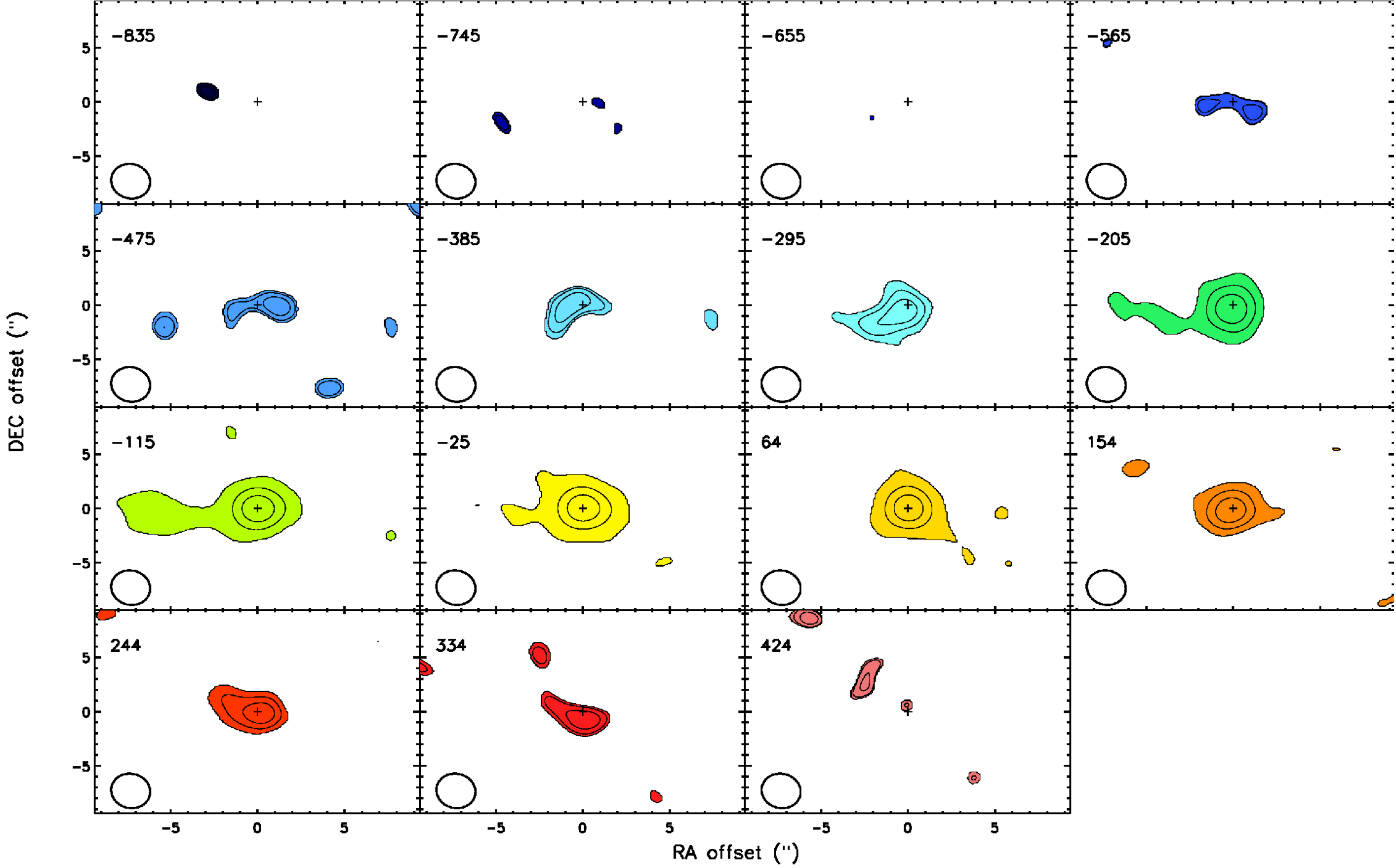}
\caption{\small{Channel map of the $^{12}$CO(1-0) line in the velocity range where the emission is detected, clipped to the 2.5$\sigma$ rms noise level. The corresponding velocity of each channel (in km~s$^{-1}$) is labelled in the top left corner. The synthesized beam size is plotted in the bottom left corner of each panel. Different colours are used to plot the emission in different channels. Contour levels are drawn at 2.5$\sigma$ intervals. R.A. and DEC offset are measured with respect to the image phase centre, highlighted by the black cross.  East  is to the left and  north is to the top.}}\label{fig:CO channel map}
\end{figure*}

The CO(J=1-0) line (115.271~GHz rest frequency) was successfully detected in Band 3. After applying the continuum self-calibration described in the previous section, the line emission was isolated by subtracting the continuum in the visibility plane using CASA's task \textit{uvcontsub}: line-free channels are identified and a first-order polynomial fit of them is performed; this continuum model is then subtracted from the data in the \textit{uv} plane to isolate the line emission. We produced the CO channel map (the  \textit{data cube}) using the CLEAN task in \textit{velocity} mode, setting a spectral resolution of $\sim90$~km~s$^{-1}$ (two channels width) and computing the velocities of each channel with respect to the rest-frame energy of the CO(1-0) line. The continuum-subtracted  dirty cube was cleaned in regions of line emission (identified interactively) to a threshold equal to 1.5 times the rms noise level determined in line-free dirty channels. The final cleaned CO data cube was produced, obtaining a 1$\sigma$ rms noise level of $\sim0.17$~mJy~beam$^{-1}$, a peak flux density of $6.14\pm0.63$~mJy~beam$^{-1}$ (significance detection=35$\sigma$), and a synthesized beam size slightly larger that that obtained for the Band 3 continuum (3.1"$\times$2.2"). The obtained CO channel map is shown in Figure~\ref{fig:CO channel map}.\\
The HCN(J=4-3) emission line (354.505~GHz rest frequency) was also detected at the 6$\sigma$ significance level. After subtracting the continuum, the HCN channel map was produced with a spectral resolution of  $\sim22$~km~s$^{-1}$. The 1$\sigma$ rms noise level measured in line-free channels is  $\sim1$~mJy~beam$^{-1}$, with a peak flux density of  $\sim6.0\pm1.1$~mJy~beam$^{-1}$. \citet{Imanishi14} presented a work including the Band 6 data of I00183 that we are analysing here. They declared the HCN emission at 3.8$\sigma$ significance level and claimed a $\sim$3$\sigma$ HCO$^{+}$(J=4-3, 356.734~GHz rest frequency) and HNC(J=4-3, 362.630~GHz rest frequency) detections. HCO$^{+}$ and HNC lines are considered undetected in our analysis, while the HCN significance is higher. These differences can be attributed to a different calibration work performed in our study with respect to that of \citet{Imanishi14}. In particular, we used a different model library release\footnote{\tiny{https://help.almascience.org/index.php?/Knowledgebase/Article/View/161/0/}}, and we reprocessed raw data, coupled with a careful editing, to improve the image dynamic range with respect to the products available from the archive. 
For completeness, we derived upper limits of the HCO$^{+}$ and HNC flux densities. Since no line emission channels can be identified even in the \textit{uv} plane, we modelled the continuum level using the spectral windows sampling continuum emission only. The continuum-subtracted data cubes were then produced with a spectral resolution of 22~km~s$^{-1}$. In the HCO$^{+}$ map we measured a 1$\sigma$ rms of 1~mJy~beam$^{-1}$, while that measured in the HNC data cube is 0.6~mJy~beam$^{-1}$. The upper limits were set assuming at least a 3$\sigma$ detection. % we estimated the HCO$^{+}$ flux density to be $<3$~mJy~beam$^{-1}$. In the HNC channel map we measured a 1$\sigma$ rms of 0.6~mJy~beam$^{-1}$ , deriving S$_{HNC<}$1.8~mJy~beam$^{-1}$.}
%It is worth noting that we obtained significant improvement in the dynamic range of the images  (e.g. from 500 to 800 in Band 6 continuum maps) with respect to the archival available products.

\begin{table*}
\begin{small}
\begin{center}
\caption{Summary of the properties of the ALMA Cycle 0 observations} \label{tab:ALMA observations summary}
\begin{tabular}{c c c c c c c c}
\hline
\hline
\multicolumn{1}{c}{\textbf{Band}} & 
\multicolumn{1}{c}{\textbf{\begin{tabular}[c]{@{}c@{}}Observation\\ date\end{tabular}}} & 
\multicolumn{1}{c}{\textbf{\begin{tabular}[c]{@{}c@{}}Sky frequency \\ coverage (GHz)\end{tabular}}} &
\multicolumn{1}{c}{\textbf{\begin{tabular}[c]{@{}c@{}}Time on target\tablefootmark{a}\\ (min)\end{tabular}}} & 
\multicolumn{1}{l}{\textbf{Flux calibrator}} &
\multicolumn{1}{c}{\textbf{\begin{tabular}[c]{@{}c@{}}PWV\tablefootmark{b}\\ (mm)\end{tabular}}} &
\multicolumn{1}{c}{\textbf{\begin{tabular}[c]{@{}c@{}}Tsys\tablefootmark{c}\\ (K)\end{tabular}}} & 
\multicolumn{1}{c}{\textbf{\begin{tabular}[c]{@{}c@{}}Number of antennas\\ (12 m)\end{tabular}}} \\ 
\hline
\hline
B3 & 18/12/2012 & \multirow{3}*{86-88/96-98} & 45.9 & Neptune &  3.4 & 70 & 25 \\

B3 & \multicolumn{1}{l}{18-19/12/2012} & & 16.1 & Callisto& 3.6 & 70 & 25 \\

B3 & 31/12/2012 & & 16.1 & Neptune& 1.9 & 60 & 24\\
\hline\hline
B6 & 29/12/2011 & \multirow{4}*{257-259/272-274} &  31.4 &    Neptune&  1.15&   100&  14\\
B6 & 13/01/2012 & &  31.4 &    Neptune&         1.6&    110&    17\\
B6 & 23/01/2012 & &   27.22 &   Neptune&        2&      110&    17\\
B6 & 23/01/2012 & &     27.22& Neptune&         2.2&    130&    17\\
\hline
B6 & 26/11/2011 & \multirow{4}*{253-255/266-268}   & 24.7 &   Callisto&                 1.5&         100&    18\\
B6 & 28/11/2011 &  &  23.8 &   Neptune&                 2&              110&         14\\
B6 & 10/01/2012 &   & 24.7 &   Neptune&                 4.2&            170&    17\\
B6 & 13/01/2012 &       & 24.7 &   Neptune&             3.3&            150&         17\\
\hline
\hline
\end{tabular}
\tablefoot{The upper panel shows the properties of Band 3 observations (unpublished data), while the two lower panels summarize the properties of Band 6 observations \citep{Imanishi14}.\\
\tablefoottext{a}{Total integration time on the science target.}
\tablefoottext{b}{Median Amount of Precipitable Water Vapour (PWV) quantified during the observation time.}
\tablefoottext{c}{System Temperature (T$_{sys}$) median value recorded in each run.}
}
\end{center}
\end{small}
\end{table*}

\section{ALMA results}\label{sec:ALMA results}
\subsection{CO as molecular gas tracer}
\begin{figure}[htbp]
\includegraphics[width=8.5cm]{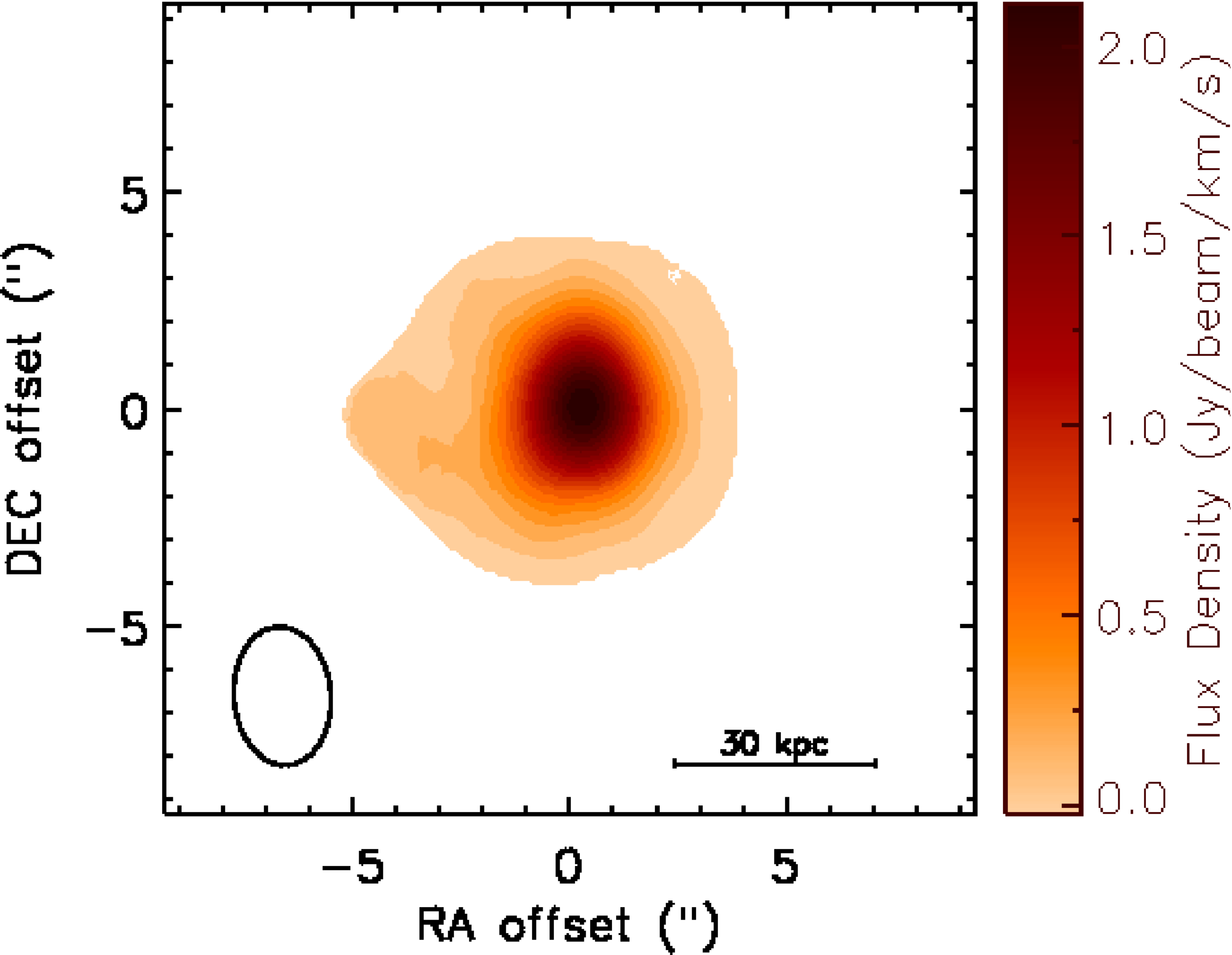}
\caption{\small{CO integrated intensity map (moment 0) created using the masked moment technique described in Section~\ref{sec:ALMA results}. The synthesized beam size is shown in the bottom left corner of the panel. The wedge on the right shows the colour-scale of the map in Jy~beam$^{-1}$~km~s$^{-1}$. After performing a 2-D Gaussian fit within the region containing the CO emission, we estimated the source size of  $(1.63\pm0.25)\times(1.09\pm0.51)$~arcsec$^{2}$ (deconvolved from beam), corresponding to $(7.8\pm1.2)\times(5.2\pm2.4)$~kpc$^{2}$. The integrated flux density is $2.56\pm0.26$~Jy~km~s$^{-1}$.}}\label{fig:CO moment 0 map}
\end{figure}

We used the cleaned continuum-subtracted CO data cube to create the final data products. The integrated intensity map (moment 0) of the detected line emission was created using the masked moment technique \citep{Dame11}. This technique consists in creating a copy of the clean data cube which is first Gaussian-smoothed spatially (with a FWHM equal to that of the synthesized beam), and then Hanning-smoothed in velocity. A 3-D mask is then defined by selecting all the pixels above a fixed flux threshold   chosen  to recover as much flux as possible in the moment map while minimizing the noise. We defined a flux threshold of 1.5$\sigma$ in this case. The moment map is then produced using the unsmoothed cubes only within the masked regions  \citep[e.g.][]{Davis17}. The obtained integrated intensity map is shown in Figure~\ref{fig:CO moment 0 map}. We also performed a 2-D Gaussian fit of the moment 0 map within the region covered by the CO emission in order to estimate its spatial extent: the molecular gas emission is resolved, with a deconvolved size of $(1.63\pm0.25)\times(1.09\pm0.51)$~arcsec$^{2}$, corresponding to $(7.8\pm1.2)\times(5.2\pm2.4)$~kpc$^{2}$ (conversion scale=4.772 kpc/", according to the adopted cosmology). The integrated line flux density is S$_{CO}=(2.56\pm0.26)$~Jy~km~s$^{-1}$ at 86.8~GHz. Following \citet{Solomon05}, we calculated the CO$_{(1-0)}$ luminosity using
\begin{eqnarray}
L'_{CO}=3.25\times10^{7}~(\dfrac{S_{CO}}{Jy~km s^{-1}})~(\dfrac{\nu_{obs}}{GHz})^{-2}~(\dfrac{D_{L}}{Mpc})^{2}~(1+z)^{-3}
,\end{eqnarray}\label{eq:luminosity}
where $\nu_{obs}$ is the observing frequency, z is the redshift, and D$_{L}$ is the luminosity distance. As a result, $L'_{CO}=1.42^{+0.08}_{-0.12}\times10^{10}$~K km s$^{-1}$ pc$^{2}$.\\
From the CO luminosity we can derive the H$_{2}$ mass using the conversion equation, expressed as in \citet{Bolatto13},
\begin{displaymath}
M(H_{2})=\alpha~L'_{CO}~M_{\odot},
\end{displaymath}
where $\alpha$ is the  H$_{2}$ mass-to-CO luminosity conversion factor. Different sources likely have different values of $\alpha$ because it is strictly dependent on the molecular gas conditions, and the environment properties (i.e. metallicity) \citep[e.g.][for a review]{Bolatto13}. In this case, we used the typical conversion factor associated with ULIRGs, $\alpha=0.8$ M$_{\odot}$~($K~km~s^{-1}~pc^{2}$)$^{-1}$ \citep{Downes98}, obtaining  M$(H_{2})=(1.14\pm0.11)\times10^{10}$~M$_{\odot}$. 
A strong empirical relation exists between the CO and the FIR luminosity and can be used to derive an estimate of the galaxy SFR. Following \citet{Carilli13},
\begin{eqnarray}
(L_\sim{IR_{starburst}})=1.37\times log(L'_{CO})-1.74 ~L_{\odot}
,\end{eqnarray}
where the IR luminosity is usually defined in the 8--1000 $\mu$m wavelengths range, and $IR_{starburst}$ refers to the fact that most of the emission which falls in this band is attributed to the starburst component. The associated SFR can be computed from
\begin{eqnarray}\label{eq:Star formation rate}
SFR\sim \delta_{MF}\times 1.0\times 10^{-10}L_{IR_{starburst}}
,\end{eqnarray}
where $\delta_{MF}$ is a factor dependent on the stellar population; for different ranges of metallicities, starburst ages, and initial mass function the value of  $\delta_{MF}$ varies in the range 0.8--2 \citep{Omont01}. Assuming a Salpeter IMF, $\delta_{MF}=1.8$ \citep{Kennicutt98}, and the SFR results to be $260\pm28$~M$_{\odot}$~yr$^{-1}$, which is comparable (within the errors) with that derived by \citet[][$\sim220$~M$_{\odot}$~yr$^{-1}$]{Mao14} from an ATCA CO$_{(1-0)}$ detection using the same method.
\subsection{Outflow hint}
\begin{figure}[h]
\includegraphics[width=9.0cm]{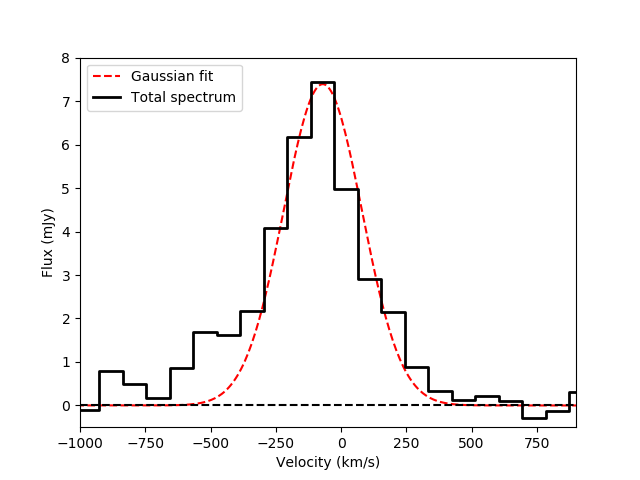}
\caption[CO spectral profiles]{\small{Spectral profiles of the CO(1-0) line emission (black) with its Gaussian fit superimposed (red). The lower axis indicates the velocity in km s$^{-1}$ relative to the line centre at $z=0.327$. The dashed horizontal line in black indicates the zero flux level. The line profile has a FWHM of $\sim400\pm27$~km~s$^{-1}$ and shows a broader and blueshifted wing, which is clearly residual with respect to the Gaussian shape. This could be considered  an indication of the presence of an outflow from the CO emitting region, but the spectral resolution is too low (90~km~s$^{-1}$) to draw solid conclusions.}}\label{fig:CO spectral profile}
\end{figure}

\begin{figure*}[h]
\includegraphics[width=9.0cm]{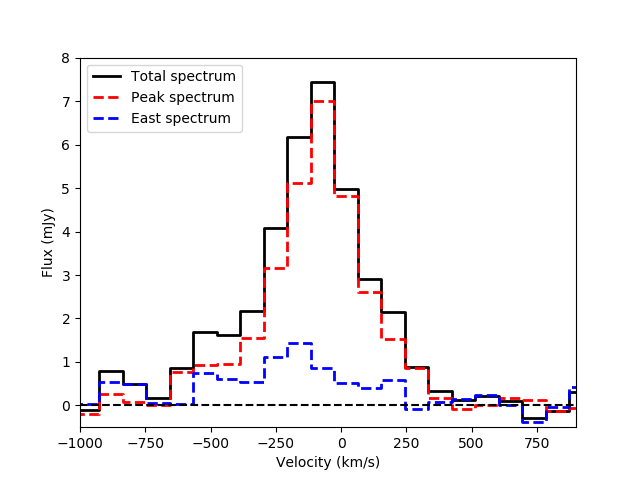}%
\includegraphics[width=7.45cm]{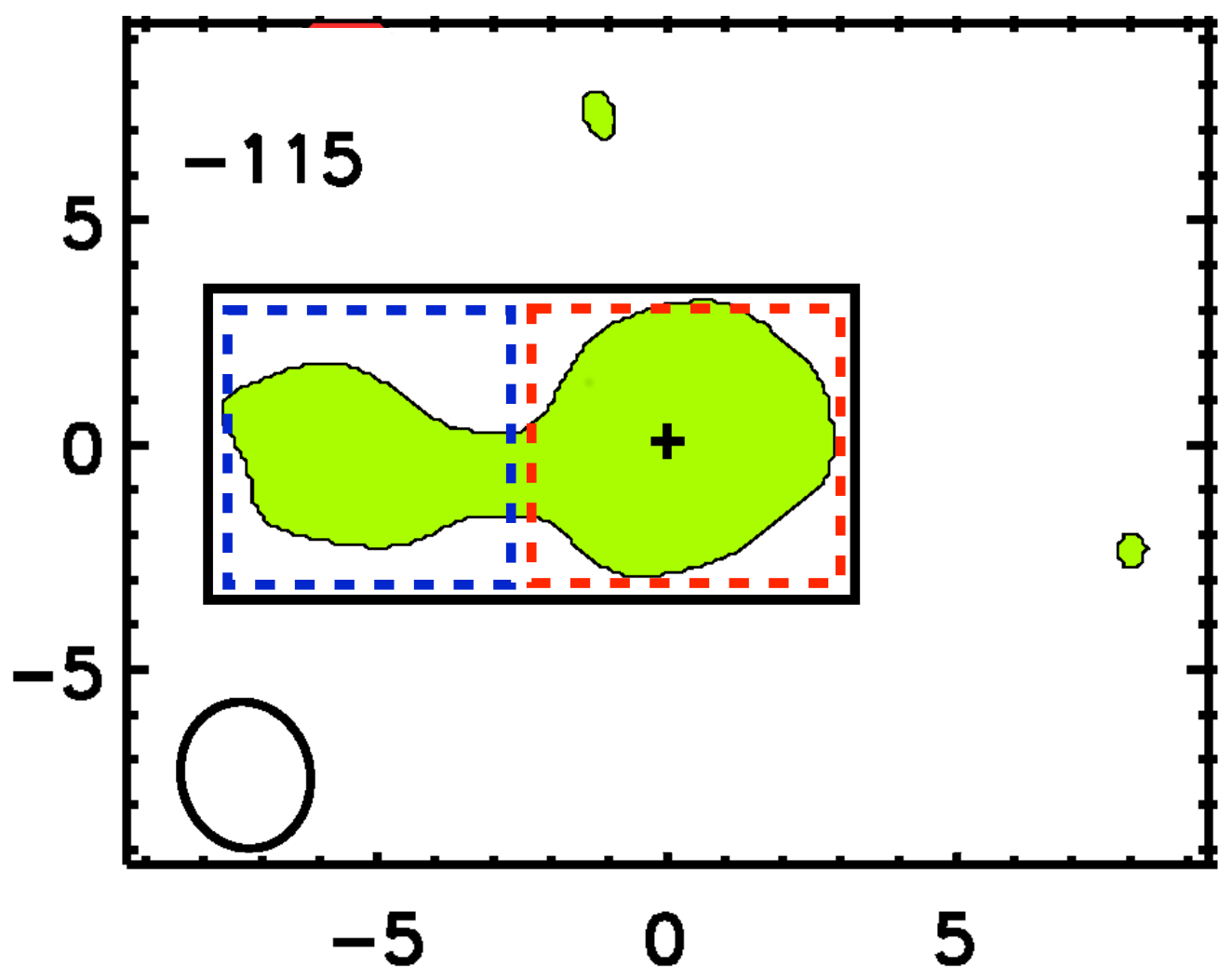}
\caption{\small{Integrated CO spectral profiles (left panel) extracted from the CO data cube within the regions highlighted in the right panel. By superimposing the three spectra it becomes clear that  the spectra around the peak and those around the eastern protrusion give their contribution to the blueshifted excess of the total integrated spectrum. Moreover, the peak of the emission extended eastward at -115~km~s$^{-1}$ (right panel) is indistinguishable in the total spectrum because of its proximity to the central line peak. Observations with higher spectral and spatial resolution would be needed to better investigate the observed features.}}\label{fig:mom0 boxes}
\end{figure*}

Figure~\ref{fig:CO spectral profile} shows the CO spectrum extracted from the channel map within a  12"$\times$8" box: we observe a Gaussian-like line profile (in black, FWHM$=400\pm27$~km~s$^{-1}$) that shows an extended, blueshifted wing which is clearly residual from the overlaid Gaussian fit (in red).
The channel map in Figure~\ref{fig:CO channel map} shows the presence of some $>$2.5$\sigma$ structures at $-$~835~km~s$^{-1}$ and $-$~745~km~s$^{-1}$, %(where the most blueshifted tail of the excess is found in the CO spectrum) 
located eastward to the nucleus; other structures around the centre are visible in the range from  -565 to -385~km~s$^{-1}$; %where the CO \textbf{spectral profile} shows the largest part of the excess. 
moreover, a protrusion to the east of the nucleus (10$\sigma$) is found at lower velocities (from -205 to -115~km~s$^{-1}$), extending up to 7" ($\sim34$~kpc) from the centre at a position angle (PA) of 87$\pm13^{\circ}$.\\
To investigate these features and assess how they contribute to the total spectrum in Figure~\ref{fig:CO spectral profile}, we extracted two spectral profiles from the boxes highlighted in the right panel of Figure~\ref{fig:mom0 boxes}: the 6"$\times$8" red and blue boxes were placed around the centre and the east protrusion, respectively; the black box shows the region from which the total spectrum (Fig.~\ref{fig:CO spectral profile}) was extracted. The left panel of Figure~\ref{fig:mom0 boxes} shows the superimposition of the three spectra. The peak of the east protrusion (at v$=-$115~km~s$^{-1}$) is adjacent to the central peak of the line (at v$=-$25~km~s$^{-1}$), and its contribution to the peak of the total spectrum is undistinguishable. The blueshifted wing in the total spectrum appears to be the result of a combination of emissions from both regions.\\
As explained in Section~\ref{sec:Source selection}, the I00183 nuclear regions host powerful high-velocity outflows inferred by line measurements of [O{\sc iii}] $\lambda$5007 \citep[][Spoon et al. in prep]{Heckman90}, [Ne{\sc ii}] 12.8$\mu$m and [Ne{\sc iii}] 15.51$\mu$m \citep{Spoon09}, OH 119$\mu$m \citep{Calderon16}, and possibly Fe{\sc xxv} \citep{Iwasawa17}. They may indicate that the nucleus is in the initial stage of disruption of the obscuring medium around the buried AGN \citep{Spoon09,Norris12}. In particular, the [O{\sc iii}] emission shows an extension in the same direction of the above-mentioned eastern protrusion. The [O{\sc iii}] image acquired from the VLT VIMOS \citep{Iwasawa17} shows an eastern extension and a bright knot at 8” (PA = 90 deg). The r-band image taken by \citet{Drake04}, which contains [O{\sc iii}] emission redshifted from 5007$\AA$, and the soft X-ray images \citep{Nandra07,Iwasawa17} show extensions towards east on the similar scale, while the K-band image of stellar light of the galaxy \citep{Rigopoulou99} shows no such an extension, indicating that the emissions extending towards the east are related only to the outflow. In addition, the much smaller scale
%As explained in section~\ref{sec:Source selection}, I00183 nuclear regions are known to host powerful high-velocity outflows of gas that likely originated from AGN wind \citep[e.g.][]{Heckman90,Spoon09,Calderon16}, and were interpreted as the initial stage of disruption of the obscuring medium around the buried AGN. The [OIII] and the 12.8$\mu$m [NeII] and 15.51$\mu$m [NeIII] outflows identified by \citet{Heckman90} and \citet{Spoon09}, respectively, show the same direction of the aforementioned East protrusion. Basing on recently acquired VLT VIMOS observations of [OIII], \citet{Iwasawa17} claimed the presence of a bright knot in the [OIII] emission at 8" east from the nucleus (PA=90$^{\circ}$, if measured from the North) and also the presence of a faint soft X-ray emission (0.4$-$1~keV) which extends up to 15" east from the nuclear point-like source (PA=70$^{\circ}$, if measured from the North). The R-band image of I00183 presented by \citet{Drake04} shows a rather broad extension up to 50~kpc east from the nucleus which should contain [OIII] emission line (redshifted from 5007$\AA$), that it is likely to be the cause of the eastward tail.
VLBI map presented by \citet[][Fig.~\ref{fig:VLBI versus ALMA}, left panel]{Norris12} shows the presence of 1.7~kpc jets that point to a similar direction to that of [O{\sc iii}] emission. From these indications, we can infer a hint of the presence of a molecular outflow from the CO emitting region. The east protrusion visible in the CO channel map (Fig.~\ref{fig:CO channel map}) is slightly tilted toward the north with respect to the [O{\sc iii}] brighter knot, but the two could be related (within the errors) and form part of a large-scale outflow. Deeper investigation of the CO line emission would be required to verify our hypothesis, but the poor spectral and spatial resolution (90~km~s$^{-1}$ and $\sim$3", respectively) and the low detection significance of the most blueshifted emission ($\sim$3.5$\sigma$) prevent us from drawing strong conclusions about this issue.  New ALMA Band 3 observations with higher spectral and spatial resolution (at least 20~km~s$^{-1}$, with a synthesized beam size of $\sim0.3$"), as well as higher sensitivity would allow us to thoroughly investigate this point.
\subsection{HCN emission analysis}
We found a HCN(4-3) emission with 6$\sigma$ significance, which is higher than that claimed by \citet[][3.8$\sigma$]{Imanishi14} using the same data, but does not allow us to carry out a detailed analysis. Line emission can be identified in only three channels of the data cube, which are not sufficient to produce a reliable integrated intensity map. We nevertheless tried to perform a 2-D Gaussian fit, measuring an integrated flux density of $\sim0.40\pm0.08$~Jy~km~s$^{-1}$, whereas the source is unresolved.\\ %An estimate of the HCN luminosity can be derived using equation~\ref{eq:luminosity} from \citet{Solomon05}, resulting that $L'_{HCN}=2.3\pm0.4\times10^{8}$~K km s$^{-1}$ pc$^{2}$.}   
The properties of the detected emission lines are summarized in Table~\ref{tab:ALMA line emission properties}.

\begin{table*}
\begin{small}
\begin{center}
\caption{\small{Properties of ALMA line detections extracted from image analysis}}\label{tab:ALMA line emission properties}
\begin{tabular}{c|ccccccc}
\hline
\hline
\multicolumn{1}{c}{} & 
\multicolumn{1}{c}{\textbf{\begin{tabular}[c]{@{}c@{}}  \textbf{$\nu_{rest}$} \tablefootmark{a} \\ (GHz) \end{tabular}}} & 
\multicolumn{1}{c}{\textbf{\begin{tabular}[c]{@{}c@{}}  $\nu_{obs}$ \tablefootmark{b}\\ (GHz)\end{tabular}}} &
\multicolumn{1}{c}{\textbf{\begin{tabular}[c]{@{}c@{}} Peak flux \tablefootmark{c}\\ (mJy~beam$^{-1}$) \end{tabular}}} & 
\multicolumn{1}{l}{\textbf{\begin{tabular}[c]{@{}c@{}}  rms \tablefootmark{d} \\ (mJy~beam$^{-1}$) \end{tabular}}} &
\multicolumn{1}{c}{\textbf{\begin{tabular}[c]{@{}c@{}} S$_{line}$ \tablefootmark{e}\\ (Jy~km~s$^{-1}$) \end{tabular}}} &
\multicolumn{1}{c}{\textbf{\begin{tabular}[c]{@{}c@{}} FWHM \tablefootmark{f}\\ (km~s$^{-1}$)\end{tabular}}} & 
\multicolumn{1}{c}{\textbf{\begin{tabular}[c]{@{}c@{}} Luminosity \tablefootmark{g}\\ (K~km~s$^{-1}$~pc$^{2}$)\end{tabular}}} \\ 
%& \textbf{$\nu_{rest}$}\tablefootmark{a} & \textbf{$\nu_{obs}$}\tablefootmark{b} & \textbf{Peak flux}\tablefootmark{c} & \textbf{rms}\tablefootmark{d} & \textbf{Flux density}\tablefootmark{e} & \textbf{FWHM}\tablefootmark{f} & \textbf{Luminosity}\tablefootmark{g}\\
\hline
CO  & 115.271 & 86.73          & $5.5\pm0.6$ ($32\sigma$)      & 0.17         & $2.56\pm0.26$         & $400\pm27$    & $1.42\times10^{10}$ \\
\hline
HCN & 354.505         & 266.72         & $5.9\pm1.1$ ($6\sigma$)    & 1.0         & $0.40\pm0.08$         & $\sim66$    & $\sim10^{9}$  \\
\hline
\hline     
\end{tabular}
\tablefoot{
\tablefoottext{a}{Rest-frame frequency of molecular lines.}
\tablefoottext{b}{Observed frequency of the line peak.}
\tablefoottext{c}{Observed line peak flux with the detection significance in parentheses.}
\tablefoottext{d}{$1\sigma$ rms noise level computed in line-free regions.}
\tablefoottext{e}{Integrated flux density of the line derived from a 2-D Gaussian fit of the emitting region.}
\tablefoottext{f}{Measured line FWHM.}
\tablefoottext{g}{Molecular line luminosity derived from the measured flux density.}
}
\end{center}
\end{small}
\end{table*}

\subsection{Continuum analysis}
\begin{figure}[h]
\centering
\includegraphics[scale=0.34]{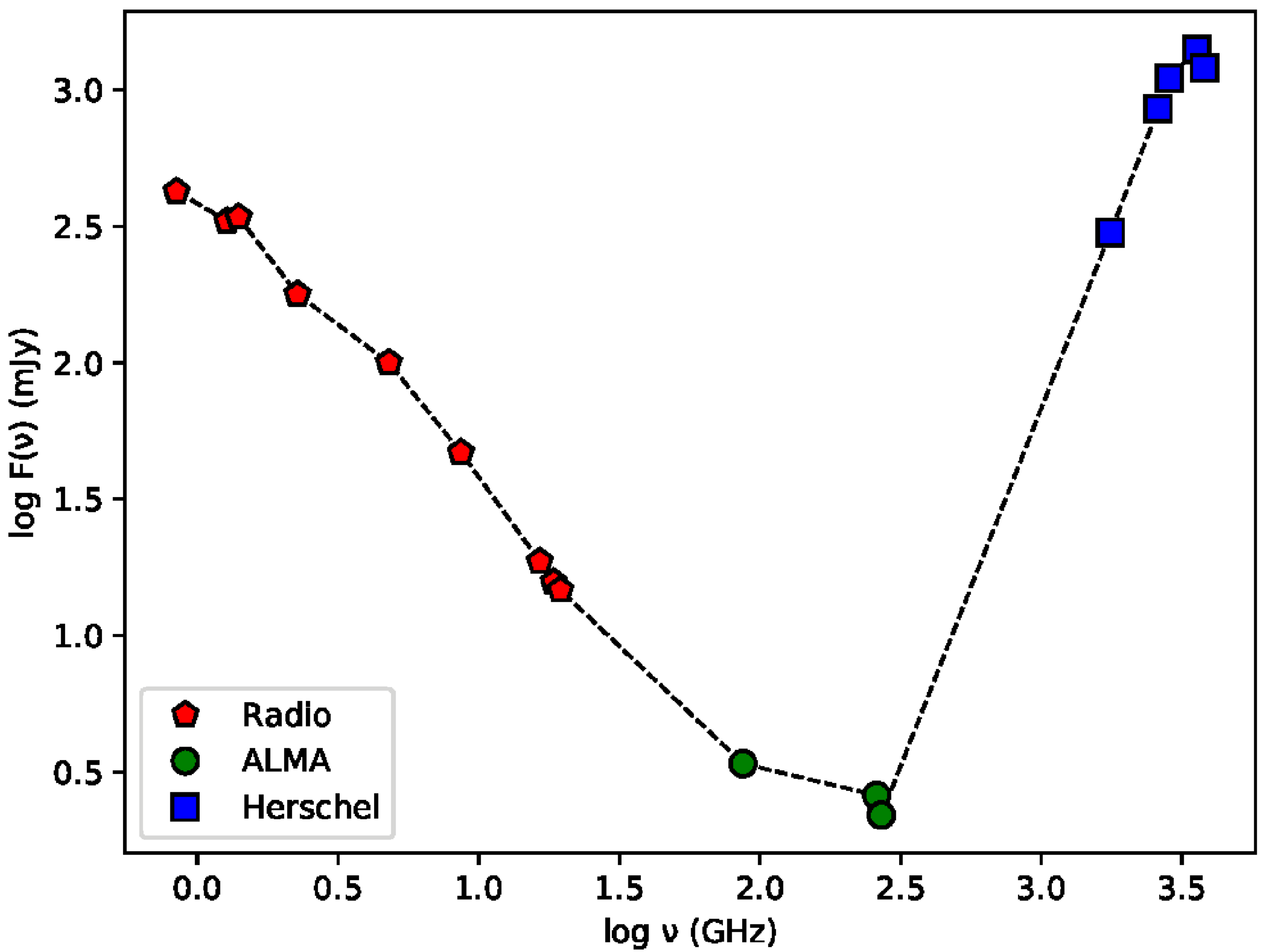}
\caption[I00183 Radio/Infrared Spectral Energy Distribution]{\small{Radio/infrared spectral energy distribution of I00183 with the observed frequencies in the lower axis. The radio, ALMA, and far-infrared flux densities are from \citet{Norris12}, this work, and the archival Herschel observations, respectively, and are indicated by red pentagons, green circles, and blue squares.}}\label{fig:I00183 Radio/IR SED}
\end{figure}

The available ALMA datasets allowed us to detect continuum emission from the nuclear regions of I00183 at different frequencies. A 2-D Gaussian fit was performed in each map within circles including all the emission; the derived properties are summarized in Table~\ref{tab:ALMA continuum emission properties}. The regions detected in the continuum are only marginally resolved, with measured integrated flux densities of 3.4$\pm0.3$, 2.6$\pm0.3$, and 2.2$\pm0.2$~mJy at 87, 263, and 270~GHz, respectively. These flux densities were then used along with radio \citep[from][]{Norris12} and Herschel (archival) data to build a radio/IR spectral energy distribution of I00183 (Fig.~\ref{fig:I00183 Radio/IR SED}; different symbols for different data). The radio data of I00183 were already analysed by \citet{Norris12}, who reported a radio spectral index (in the range 0.84$-$87~GHz) $\alpha=-1.38$ (where $S_{\nu}\propto \nu^{\alpha}$), likely dominated by synchrotron emission from the powerful nuclear AGN activity in I00183 \citep{Roy97,Spoon09}. The ALMA flux densities yield a flat spectral index, $\alpha=-0.24\pm0.10$ between 87 and 270~GHz. % the 270~GHz flux density was not considered here because is consistent (within the errors) with the 263~GHz measurement. 
Thermal emission from dust is expected to give its contribution to the continuum emission at millimetre wavelengths, becoming likely the dominant emission mechanism above 200~GHz ($\alpha_{th} \sim 3-4$, \citealp[see e.g.][]{Condon92,Scoville14}).\\ % ($S_{\nu}\propto \nu^{\alpha}$).
To quantify the contribution of thermal emission from dust in the 87$-$270~GHz range, we fit the FIR part of the SED in Figure~\ref{fig:I00183 Radio/IR SED} (based on the archival Herschel flux densities) with a single-temperature modified black-body function (the so-called `grey body') in the form
 \begin{eqnarray}\label{eq:dust}
S(\nu_{0})= \dfrac{M_{d}~(1+z)~k(\nu_{r})~B(\nu_{r},T_{d})}{D_{L}^{2}}
,\end{eqnarray}
where S$(\nu_{o})$ is the observed flux density, $M_{dust}$ is the dust mass, $B(\nu_{r},T_{d})$ is the Planck function at the rest-frame frequencies $\nu_{r}$,  D$_{L}$ is the luminosity distance, and $k(\nu_{r})$ represents the frequency-dependent dust absorption coefficient in [m$^{2}$~kg$^{-1}$] described by a power law with dust emissivity index $\beta$. For our purpose, we assumed the value of k$_{o}$ tabulated by \citet[][Table 6]{Li01} at the reference observed frequency $\nu_{o}=2141$~GHz ($\lambda_{o}=$140~$\mu$m) and applied $k(\nu_{r})=k_{o}(\nu_{r}/\nu_{o})^{\beta}$ \citep[e.g.][]{Magdis11,Magnelli12}. The value of $\beta$ is typically poorly known, being strongly dependent on the dust grain compositions, temperatures, and sizes. However, it was found to vary between 1.5 and 2.0 within a wide range of infrared dust emissions in strong millimetre-emitting sources \citep[e.g.][]{Magnelli12}. We fixed the value of the emissivity index to 1.5. The dust temperature (T$_{d}$) and mass ($M_{d}$) were left free to vary in the fit. We obtained $T_{d}=68\pm10$~K and $M_{d}=(2.7\pm1.4)\times10^{7}$~M$_{\odot}$ as best-fit parameters; the errors were quoted assuming a 20\% flux density uncertainty in the archival Herschel measurements. The best-fit dust temperature is higher than the typical values found in ULIRGs \citep[e.g.][]{Farrah03}, although emission from a `warm' ($>40$~K) dust component is compatible with the presence of vigorous star formation and AGN activities in the few central kiloparsec of this source \citep[e.g.][]{Lisenfeld00,Spoon09}. From our analysis, the best-fit dust mass is lower than that typically found in local ULIRGs \citep[e.g.][]{Farrah03}. The simplistic approach of a single-temperature modified black-body model, the lack of sufficient information in the optically thin dust regime (i.e. above $\sim$200~$\mu$m), and most importantly the assumption we made on $\beta$ may lead us to bias the obtained results in this direction. However, our main goal is to estimate the contribution of dust emission at the frequencies of the ALMA observations (87$-$270~GHz);  assuming $\beta=$1.5 allows us to put a conservative upper limit to that contribution. Using the best-fit parameters and applying equation~\ref{eq:dust}, we found that thermal emission from dust would contribute  less than 4\% at 87~GHz (3.3~mm), while it is the dominant emission mechanism at 270~GHz (1.3~mm). As a consequence, the emission at 87~GHz would be consistent with the tail of radio synchrotron emission, while that at 270~GHz would be consistent with the Rayleigh--Jeans tail of thermal emission from dust. The flat spectral index measured between 87 and 270~GHz is likely attributable to a mixture of optically thin synchrotron emission and thermal dust emission. In addition, the vigorous starburst hosted in the central regions of I00183 \citep[SFR$\sim$260~M$_{\odot}$~yr$^{-1}$][this work]{Nandra07,Mao14} heats and photoionizes the surroundings, then free-free emission from ionized atomic gas may also contribute to the millimetre continuum spectral slope ($\alpha_{FF}$ is expected to be in the range from $-$0.1 to 0.4 at millimetre/sub-millimetre wavelengths; \citealp[see e.g.][]{Pascucci12}). Additional ALMA continuum measurements (100--900~GHz) would enable us to carry out a detailed study of the millimetre/sub-millimetre continuum emission of I00183 and confidently establish the dominant emission mechanism in the few central kiloparsec of this source. 
\subsection{Line-continuum emission offset}
\begin{figure*}[htbp]
\centering
\includegraphics[scale=0.6]{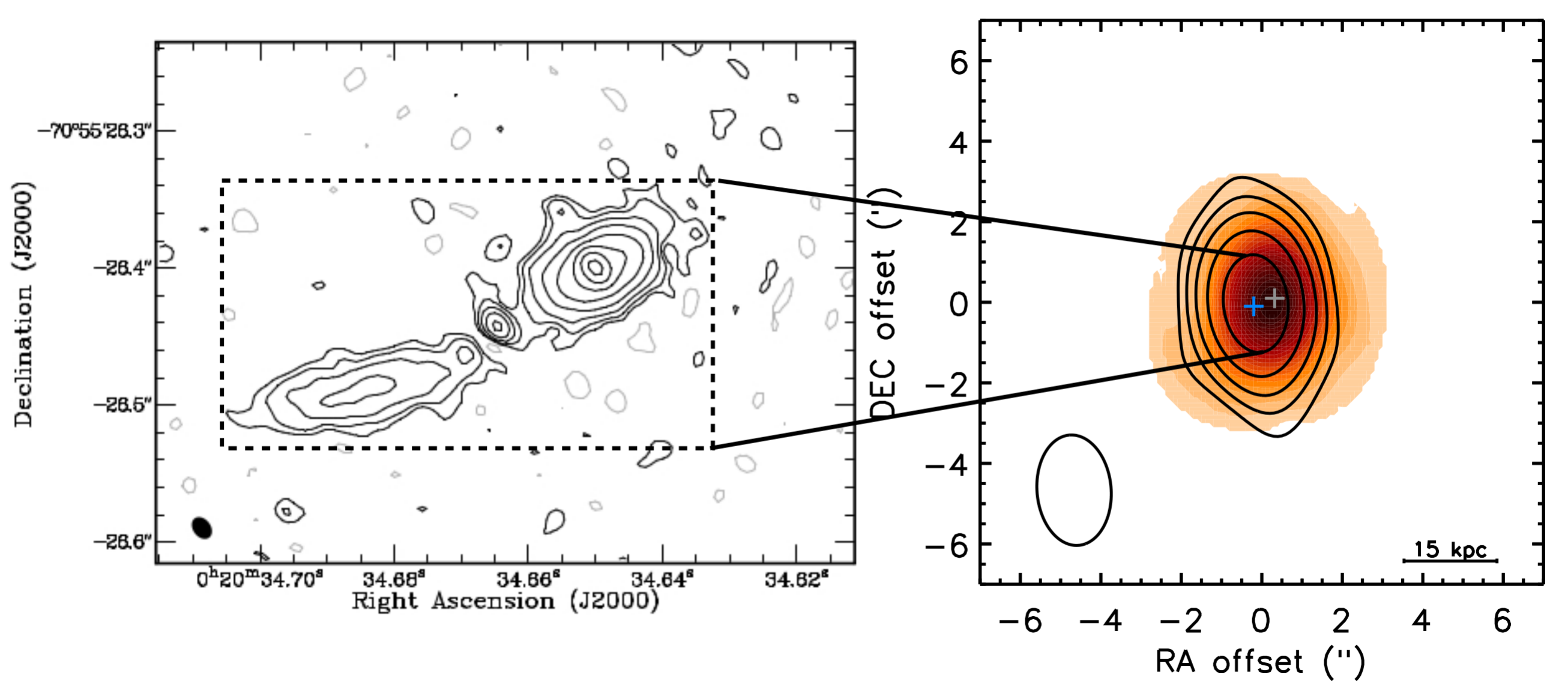}
\caption{\small{VLBI map of I00183 at 2.3~GHz (left panel) adapted from \citet{Norris12} and ALMA CO integrated intensity (moment 0) with the underlying 87~GHz continuum contours superimposed (right panel). Contours are drawn at 3,6,9, $...$ times the $1\sigma$ rms noise level of the continuum map ($0.02$~mJy~$^{-1}$). The white and blue crosses indicate the position of the centre of ALMA CO and continuum emissions, respectively; the two positions are shifted by 0.53 arcsec$\sim2.5$~kpc. The spatial extent of the region imaged by the VLBI observation is completely embedded into the ALMA continuum synthesized beam (shown in the bottom left corner of the right panel), but the orientation of the radio source seems to be consistent with the direction of the observed shift between the line and the continuum emission in ALMA Band 3.}}\label{fig:VLBI versus ALMA}
\end{figure*}

By superimposing the 87~GHz continuum and CO moment 0 maps, we found a shift between the line and the continuum emission peaks (Fig.~\ref{fig:VLBI versus ALMA}, right panel). The coordinates of the two positions were obtained by performing 2-D Gaussian fits within circles around the peaks. The error bar associated with  the shift was derived propagating the astrometric errors only since the peak position errors obtained from the image Gaussian fitting were around a factor of 4 smaller. The astrometric uncertainties of the two positions in each map were estimated as $\sigma_{p}\sim beamsize/S/N$ \footnote{https://almascience.nrao.edu/documents-and-tools/cycle5/alma-technical-handbook/view, pag.~154.}. The observed offset is 0.53"$\pm$0.08" ($\sim7 \sigma$), corresponding to $2.5\pm0.4$~kpc. It is also worth noting that the measurements were  performed using the Briggs weighted maps (robust=0.5), with resolutions of 2.57"$\times$1.69" and 2.74"$\times$1.84" for continuum and line images, respectively.\\
It is challenging to give a solid interpretation of this shift, mainly because the spatial resolution of Band 3 observations does not allow us to resolve both the line and the continuum emissions. However, on a purely speculative basis, we can attempt to find a possible explanation. It is already well established that I00183 hosts a powerful radio source in its central regions with $L_{4.8 GHz}=3\times10^{25}$~W~Hz$^{-1}$ \citep{Roy97}, which places it within the regime of the FRII (i.e. high-luminosity) radio galaxies. Furthermore, VLBI observations of I00183 at 2.3~GHz presented by \citet{Norris12} show the presence of a compact radio-loud AGN in the galaxy nuclear regions; it presents quasar jets only 1.7~kpc long, boring through the dense gas and dust layers that still confine them. \citet{Norris12} argue that the morphology and the spectral index of the source are consistent with compact steep spectrum (CSS) sources, which are widely thought to represent an early stage of evolution of radio galaxies \citep[e.g.][]{Odea98}: it seems plausible that I00183 AGN has just switched on and started boring its way through the nuclear environment. Against this backdrop, the shift observed between the line and the continuum maps in Band 3 may be interpreted as the result of an interaction between the `newborn' radio source and the surrounding gas; interestingly, the orientation of the radio source observed by \citet{Norris12} seems to be consistent with the direction of the shift (Fig.~\ref{fig:VLBI versus ALMA}) and, at the same time, the 87~GHz continuum seems to peak eastward where the western jet visible in the VLBI map is stronger. However, the AGN structure that is resolved in detail in the VLBI map by \citet{Norris12} is instead completely embedded into the Band 3 ALMA beam size (Fig.~\ref{fig:VLBI versus ALMA}), and the spatial distribution of the molecular gas is only marginally resolved, which  prevents us from drawing strong conclusions. New ALMA observations of I00183 with higher spatial (and spectral) resolution would allow us to resolve the inner regions of I00183 and better investigate this point.

\begin{table*}
\begin{small}
\begin{center}
\caption{\small{ALMA continuum emission properties}} \label{tab:ALMA continuum emission properties}
\begin{tabular}{l|cccccc}
\hline
\hline
\multicolumn{1}{c}{} & 
\multicolumn{1}{c}{\textbf{\begin{tabular}[c]{@{}c@{}}  Freq. ranges \tablefootmark{a} \\ (GHz) \end{tabular}}} & 
\multicolumn{1}{c}{\textbf{\begin{tabular}[c]{@{}c@{}}  Flux density \tablefootmark{b}\\ (mJy)\end{tabular}}} &
\multicolumn{1}{c}{\textbf{\begin{tabular}[c]{@{}c@{}} Rms  \tablefootmark{c}\\ (mJy~beam$^{-1}$) \end{tabular}}} & 
\multicolumn{1}{l}{\textbf{\begin{tabular}[c]{@{}c@{}}  Beam size \tablefootmark{d} \\ (") \end{tabular}}} &
\multicolumn{1}{c}{\textbf{\begin{tabular}[c]{@{}c@{}} Source size  \tablefootmark{e}\\ (kpc$^{2}$) \end{tabular}}} &
\multicolumn{1}{c}{\textbf{\begin{tabular}[c]{@{}c@{}} Luminosity \tablefootmark{f}\\ (erg~s$^{-1}$)\end{tabular}}}\\
%\multicolumn{1}{c}{\textbf{\begin{tabular}[c]{@{}c@{}} Luminosity \tablefootmark{g}\\ (K~km~s$^{-1}$~pc$^{2}$)\end{tabular}}} \\ 
%& \textbf{Freq. ranges}\tablefootmark{a} & \textbf{Flux}\tablefootmark{b} & \textbf{rms}\tablefootmark{c} & \textbf{Beam pattern}\tablefootmark{d} &\textbf{Source size}\tablefootmark{e} & \textbf{Luminosity}\tablefootmark{f} \\
\hline
Band 3 & 86.1-87.8/96.2-97.7 & $3.4\pm0.3$ & 0.016 & $3.01\times2.05$ & $(3.37\pm0.2)\times(2.26\pm0.1)$ & $\sim8.2\times10^{41}$ \\
\hline
\multirow{2}{*}{Band 6} & 257.2-259.11/272.35-274.22 & $2.2\pm0.2$ & 0.015 & $2.35\times1.19$ & $(4.8\pm1.1)\times(3.8\pm1)$ & $\sim2.1\times10^{42}$ \\
& 253.3-255.2/266.2-267.8 & $2.6\pm0.3$ & 0.05 & $2.48\times1.31$ & $(3.9\pm0.8)\times(2.7\pm0.52)$ & $\sim2.4\times10^{42}$ \\
\hline
\hline                      
\end{tabular}
\tablefoot{
\tablefoottext{a}{Observed frequency ranges.}
\tablefoottext{b}{Integrated continuum flux density.}
\tablefoottext{c}{$1\sigma$ rms noise level, computed in source-free regions.}
\tablefoottext{d}{Synthesized beam sizes of natural weighted continuum maps.}
\tablefoottext{e}{Source spatial extent deconvolved from the clean beam.}
\tablefoottext{f}{Luminosity associated with the continuum emission in each band.}
}
\end{center}
\end{small}
\end{table*}
  
\section{X-ray observations and spectral fitting}\label{X-ray obs}
The X-ray properties of I00183 were studied using archival \chandra\ and \xmm\ data, while \nus\ data were obtained by courtesy of the PI (K. Iwasawa; see \citealp{Iwasawa17} for details). The main properties of X-ray observations are summarized in Table~\ref{tab:X-ray observation summary}. \chandra\ and \xmm\  data were reprocessed, and reduced spectra were analysed using XSPEC software, version 12.9.0 (Arnaud et al. 2015).\\
We carried out the X-ray analysis of I00183 first by analysing individually \chandra, XMM, and \nus\ spectra. We started the spectral fitting using simple phenomenological models and then introducing more complex ones. All the data were finally used for the broad-band joint spectral analysis once we realized that a similar model provides a good representation of the individual datasets. In the following, the errors are reported at the 90\% confidence level for one parameter of interest \citep{Avni76}. 

\subsection{Chandra data}
The 22.8~ks \chandra\ observation was taken on 13 February 2013 using the ACIS-S detector. The raw data were reprocessed using CIAO software (version 4.7). A region of 2~arcsec was chosen to extract the target spectrum, while a larger nearby region, free of contaminating sources, was selected for the background. The \textit{specextract} tool was finally run to create the spectra, along with the ARF and RMF response matrices. Due to the limited number of spectral counts (133), we chose to apply the Cash statistics, which is more appropriate in low-count regimes, and we applied a minimum binning of one count per bin to avoid empty channels \citep[see][Appendix A]{Lanzuisi13}. The net count rate for the spectrum is $5.8\times10^{-3}~cts~s^{-1}$, 99.6\% of which are source counts.
\subsubsection{Chandra spectral analysis}
At the beginning, a simple spectral model composed of a power law modified by the Galactic absorption\footnote{The Galactic absorption was calculated using the web tool available at  http://heasarc.nasa.gov/cgi-bin/Tools/w3nh/w3nh.pl. Throughout all the spectral analyses we adopted a Galactic absorption value of $3.27\times10^{20}~cm^{-2}$ taken from the Leiden Argentine Bonn (LAB) Survey of Galactic HI.} (\textit{wabs}) was adopted. The fit (c-stat/dof$=99/108$) leaves significant residuals, suggesting that  more complex modelling is needed to obtain a good agreement between the data and the model. 
We find $\Gamma=-0.48^{+0.27}_{-0.28}$, which is much flatter than the photon index expected in 
unobscured AGN ($\Gamma=$1.8--1.9; e.g. \citealt{Piconcelli05}) and is 
suggestive of the presence of intrinsic absorption. Consequently, we added a component to model the absorption at the source redshift (\textit{zwabs}), and a narrow (10~eV, fixed) line component (\textit{zgauss}) to model the clearly present iron emission line. The resulting intrinsic column density N$_{H}$ is $5.3^{+3.8}_{-2.8}\times10^{22}$~cm$^{-2}$ with an observed photon index $\Gamma=1.3^{+0.9}_{-0.8}$; the line centroid is at $6.77\pm0.08$~keV, which is consistent with the rest-frame energy of the iron helium-like K$\alpha$ line, as already found in \citet{Nandra07}. The line has a rest-frame equivalent width (EW) of $0.5^{+0.4}_{-0.3}$~keV, and the overall improvement with respect to the previous modelling is evident (c-stat/dof$=78/106$). 
Fixing $\Gamma=1.9$ would imply a column density that is  higher by $\sim$50\%. \\
%Even if the phenomenological model produces a good fit, we adopted \textsc{MYTorus} \citep{Murphy09} model to possibly provide a more physical characterization of the source emission. \textsc{MYTorus} takes into account the geometry of the obscuring medium and consists of three component: (1) intrinsic absorption, which modifies the primary power law emission, (2) reflection component to model the Compton reflection continuum, (3) line components. However, component (3) models only the 6.4 K$\alpha$ and 7.1 K$\beta$ iron lines; thus, in this particular case, we modelled the 6.7 keV Helium-like iron line using an additional single Gaussian component. The inclination angle of the source with respect the line of sight was found to be $\sim70^{\circ}$ and then frozen. 
Given the presence of an ionized He-like emission line, we added to the absorption model described above a reflection component arising from ionized gas. In particular, we adopted the \textsc{Xillver} model \citep{Garcia10,Garcia13}, which  ensures a proper treatment of the ionized reflection and provides an accurate description of the Fe K lines. The ionization state of the medium is described by the ionization parameter, defined as $\xi=4\pi F_{x}/n$, where F is the flux of the illuminating radiation in the $\sim$0.1$-$100 keV energy range, and $n=10^{15}$~cm$^{-3}$ is the gas density. %The ionization parameter was left free to vary and its best-fit value was found to be log$\xi={2.35\pm0.04}$, indicating a medium in a strongly ionized state. 
We note that the spectrum also shows a soft component (below 2 keV) that is probably due to either thermal emission from the host galaxy or a scattering component (or a mixture of the two); this was modelled using a simple power law since the photon statistics are limited. We also added a Gaussian line component to account for the neutral iron line emission at 6.4 keV (fixed), which is expected in the case of cold transmission \citep[e.g.][]{Leahy93}; its normalization is fixed at EW$\sim$100~eV (rest frame). The resulting best-fit (c-stat/dof$=79/106$) intrinsic column density is $N_{H}=(9\pm2 )\times10^{22}$~cm$^{-2}$, with the photon index fixed at the expected value of 1.9 to provide more stringent constraints to the intrinsic absorption, given the degeneracy between $\Gamma$ and $N_{H}$. The inclination angle of the ionized reflection component is not well constrained because of the limited statistics, thus it was initially left free to vary and then fixed at its hard limit of 87$^{\circ}$. The ionization parameter is log$\xi=3.04^{+0.61}_{-0.22}$, consistent with that expected when an ionized He-like iron emission line is detected \citep[e.g.][]{Gilli14, Iwasawa17}. As expected, the best-fit relative strength between the reflection and the power law components is about $10^{-3}$, which means that the reflection component is fainter than the primary power law emission.  
%
%We note that a similar value for the neutral absorption is found if the \textsc{MYTorus}\footnote{\textsc{MYTorus} takes into account the geometry of the obscuring medium and consists of three component: (1) intrinsic absorption, which modifies the primary power law emission, (2) reflection component to model the Compton reflection continuum, (3) line components.} \citep{Murphy09} model is adopted to provide a more ``physical" characterization of the source emission. 
%
The observed best-fit model flux is $(1.46^{+0.26}_{-0.23})\times10^{-13}$~erg~s$^{-1}$~cm$^{-2}$ in the observed energy range 2--10 keV, while the de-absorbed (i.e. corrected for  absorption) 2--10 keV rest-frame luminosity is $(7.4^{+1.3}_{-1.2})\times10^{43}$~erg~s$^{-1}$.
%The obtained spectral parameters seems to converge to the idea that a strongly obscured AGN source is embedded into the nuclear region of I00183, whose primary power law emission is flattened by the transmission through the obscuring medium that intercept the line of sight, while the reflection component arises from ionized gas, with iron in the He-like state. %The obtained best-fit spectral fitting is shown in Figure~\ref{fig:Chandra My Torus}. 
%We stress that \textit{Chandra} data were useful to better examine the X-ray properties of the source, but the spectrum was excluded from the broad-band spectral fitting along with XMM and \textit{NuSTAR} because of its lower statistics.

\subsection{XMM-\textit{Newton} data}
The galaxy I00183 was observed by \xmm\ on 16 April 2013 with a nominal integration time of 22.2~ks. We started the analysis from the ODF raw data format and reprocessed them using the Science Analysis System (SAS) software, version 14.0. Data reduction was carried out separately for the pn and MOS1-MOS2 cameras using the standard \textit{epproc} and \textit{emproc} tasks, respectively. We produced a light curve at energies 10--15~keV to inspect for possible periods of background flares and select good time intervals (gti); the observation appears highly flared, thus we adopted the following screening in selecting gti, aiming to achieve the best compromise between the S/N and the effective exposure time: CR$<5$~counts~s$^{-1}$ for the pn, and $<2$~counts~s$^{-1}$ for the two MOS cameras. The remaining useful exposure times for the EPIC pn and MOS cameras are 12.0~ks and $8.4+8.4$~ks, respectively. The pn spectrum was extracted within a region of 15 arcsec of radius, while the MOS spectra were extracted from a region of 20 arcsec of radius; both correspond to an encircled energy fraction of $\sim75\%$ at 1.5~keV. We  tested different spectral binning and, finally, a good compromise between the S/N and the spectral data counts was achieved by grouping the pn spectrum by 15 counts for bin (290 spectral data counts), and the co-added MOS1-MOS2 spectra by 10 counts for bin (170 spectral data counts); the $\chi^2$ statistics was adopted. The net count rate is $1.3\times10^{-2}$~counts~s$^{-1}$ (57.4\% from the source), and $6.3\times10^{-3}$~counts~s$^{-1}$ (63\% from the source) in the pn and MOS1-MOS2 cameras, respectively.
\subsubsection{XMM-Newton spectral fitting}
The \xmm\ spectra suggest an X-ray modelling similar to that outlined by \chandra, i.e. an absorbed power law plus a soft component, and an  ionized reflection; we also included a constant to account for the cross-normalization of the different instruments 
%(pn and two MOS cameras, consistent within 80\%) 
and a Gaussian component at the fixed energy of 6.4 keV to account for the neutral iron emission line expected in the case of transmission through a cold medium (EW$\sim$100 eV). We obtained a best-fit ($\chi^{2}$/dof$=40/38$)  $N_{H}=(6.8^{+3.0}_{-1.8})\times10^{22}$~cm$^{-2}$, with $\Gamma$ fixed at 1.9; the ionization parameter is log$\xi=2.68^{+0.36}_{-0.92}$. The observed-frame 2--10~keV flux is $(1.35^{+0.61}_{-0.32})\times10^{-13}$~erg~s$^{-1}$ cm$^{-2}$, which is consistent with that measured using the \chandra\ data. The de-absorbed intrinsic luminosity in the 2--10~keV range is $(5.6^{+2.6}_{-1.3})\times10^{43}$~erg~s$^{-1}$. 

\subsection{\textit{NuSTAR} data}
The galaxy I00183 was observed by \nus\ twice between 21 December 2015 and 26 April 2016 for a total exposure time of 105~ks \citep{Iwasawa17}. Four spectra were extracted from the observations of the two detector modules FPMA and FPMB, then the two FPMA and the two FPMB spectra were combined. We grouped the spectra by 15 counts per bin (spectral data counts of 240, 220 and 270, 200, respectively), and used the $\chi^{2}$ statistics. The spectral analysis was restricted within the observed 3--24~keV energy range. The net photon count rate is $\sim3\times10^{-3}$~counts~s$^{-1}$ for the four spectra; about 60\% of these counts are from the source.

\subsubsection{\textit{NuSTAR} spectral fitting}
Starting from the previous modellings of \chandra\ and \xmm\ spectra, we used the \textsc{Xillver} model to account for the reflection component, a simple intrinsic cold absorption component, and a Gaussian line to model the iron emission at 6.4 keV due to transmission; we also added a constant to account for the cross-normalization between the two focal plane modules.
The cold gas column density cannot be well constrained in this case due to the limited \nus\ low-energy coverage, thus we obtained an upper limit for the obscuration of $<10^{23}$~cm$^{-2}$ (with $\Gamma$ fixed to 1.9) and limited residuals ($\chi^{2}/dof=37/46$). Although the ionized iron emission is much less evident than in the previously fitted spectra (see also the discussion in \citealp{Iwasawa17}  in this regard), the best-fit ionization parameter is within the range of \textit{Chandra} and \xmm\ data (log$\xi=3.2^{+0.4}_{-0.3}$). The observed-frame 2--10~keV flux is $(1.7^{+0.1}_{-0.4})\times10^{-13}$~erg~cm$^{2}$~s$^{-1}$, which is consistent (within the errors) with the measurements reported above; the rest-frame 3--24~keV luminosity is $\sim(1.3^{+0.2}_{-0.2})\times10^{44}$~erg~$s^{-1}$.

\begin{table*}
\caption{Summary of the X-ray observations} \label{tab:X-ray observation summary}
\begin{center}
%\begin{scriptsize}
\begin{tabular}{ l | l l l l l l l}
\hline
\hline
\textbf{Satellite} & \multicolumn{1}{c}{\textbf{\begin{tabular}[c]{@{}c@{}}  Exposure time \\ (ks) \end{tabular}}} & 
\multicolumn{1}{c}{\textbf{Observation date}} & 
\multicolumn{1}{c}{\textbf{ObsID}}&
\multicolumn{1}{c}{\textbf{Instrument}}&
\multicolumn{1}{c}{\textbf{\begin{tabular}[c]{@{}c@{}}  Band\\ (keV) \end{tabular}}}\\
\hline
\textit{Chandra} & \multicolumn{1}{c}{22.8} & \multicolumn{1}{c}{13/02/2013} & \multicolumn{1}{c}{13919} &\multicolumn{1}{c}{ACIS-S} & \multicolumn{1}{c}{0.5$-$7.2}\\
\hline
XMM-\textit{Newton} & \multicolumn{1}{c}{22.2} & \multicolumn{1}{c}{16/04/2003} & \multicolumn{1}{c}{0147570101}&\multicolumn{1}{c}{EPIC pn-MOS 1/2}& \multicolumn{1}{c}{0.5$-$7.2}\\
\hline
\textit{NuSTAR} & \multicolumn{1}{c}{115} & \multicolumn{1}{c}{21/12/2015--26/04/2016} & \multicolumn{1}{c}{6010105002/4} &\multicolumn{1}{c}{FPMA-FPMB} & \multicolumn{1}{c}{ 3$-$24}\\
\hline
\hline
\end{tabular}
%\end{scriptsize}
\end{center}
\end{table*}

\begin{table*}[htbp]
\caption{\small{Best-fit X-ray parameters obtained using the  \textsc{Xillver} model}} \label{tab:X-ray MyTorus parameters}
\begin{center}
\begin{tabular}{ l | l l l l l l l}
\hline
\hline
&\multicolumn{1}{c}{\textbf{\textbf{log$\xi$}\tablefootmark{a}}} & 
\multicolumn{1}{c}{\textbf{$\Gamma$}\tablefootmark{b}} & 
\multicolumn{1}{c}{\textbf{N$_{H}$\tablefootmark{c}}} & 
\multicolumn{1}{c}{\textbf{Flux\tablefootmark{d}}} & 
\multicolumn{1}{c}{\textbf{Luminosity\tablefootmark{e}}} \\
\hline
\textit{Chandra} & 3.04$^{+0.61}_{-0.22}$ & 1.9 (fixed) & $9.0\pm2.0$ & 1.46$^{+0.26}_{-0.23}$ & 7.46$^{+1.29}_{-1.15}$\\
\hline
XMM-\textit{Newton} & $2.68^{+0.36}_{-0.92}$ &  1.9 (fixed) & $6.8^{+3.0}_{-1.8}$ & 1.35$^{+0.61}_{-0.32}$ & 5.65$^{+2.57}_{-1.33}$ \\
\hline
\textit{NuSTAR} &  $3.2\pm0.3$ & 1.9 (fixed) & $<10$ & $1.7^{+0.4}_{-0.1}$ & 8.1$^{+1.43}_{-1.35}$\\
\hline
%\hline
%& \multicolumn{1}{c}{\textbf{log$\xi$}\tablefootmark{f}} & \multicolumn{1}{c}{\textbf{$\Gamma$}\tablefootmark{g}} & \multicolumn{1}{c}{\textbf{N$_{H}$}\tablefootmark{c}} & \multicolumn{1}{c}{\textbf{Flux}\tablefootmark{d}} & \multicolumn{1}{c}{\textbf{Luminosity}\tablefootmark{e}}\\
%\hline

\hline
\multicolumn{6}{c}{\small{\textbf{Broad-band spectral parameters (0.5$-$24 keV)}}}\\
\hline
\multicolumn{1}{l}{}&  $2.6\pm0.15$ &  1.9 (fixed) & $6.8^{+2.1}_{-1.5}$ & $1.44^{+0.35}_{-0.37}$ &  $4.2\pm0.1$\\                         
\hline
\hline
\end{tabular}
\tablefoot{
\tablefoottext{a}{Ionization parameter derived from the best fit, with its errors quoted at the 90\% confidence level.}
\tablefoottext{b}{Photon index derived from the best fit with errors quoted at the 90\% confidence level.}
\tablefoottext{c}{Intrinsic N$_{H}$ column density in units of $10^{22}$~cm$^{-2}$ and obtained from the best-fit model.}
\tablefoottext{d}{Observed-frame 2--10 keV flux from the best spectral fitting in units of $10^{-13}$ erg $cm^{-2}$ s$^{-1}$.}
\tablefoottext{e}{Intrinsic rest-frame 2--10 keV luminosity from the best spectral fitting in units of $10^{43}$ erg s$^{-1}$.}
}
\end{center}
\end{table*}

\subsection{Broad-band spectral fitting}
\begin{figure}[h]
\centering
\includegraphics[width=10cm]{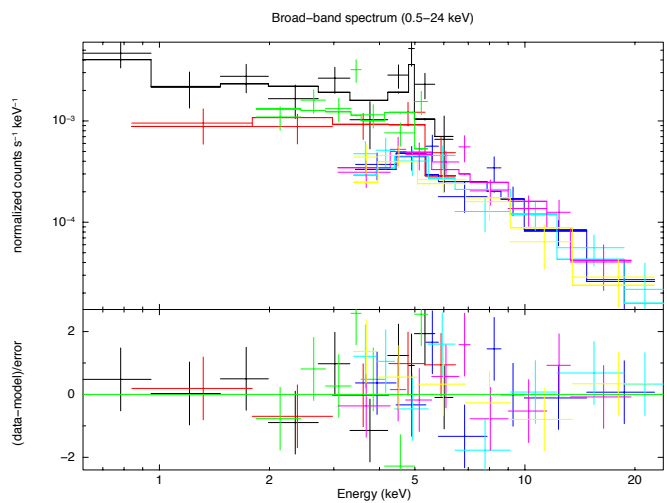}
\caption[Broad band spectral fitting using \textsc{Xillver} model]{\small{I00183 \textit{Chandra}, \xmm, and \textit{NuSTAR} spectra folded through the instrumental response with the model superimposed in the observed energy frame 0.5$-$24~keV. All the spectra are grouped by 15 counts per bin and further re-binned for presentation purposes. The model consists of \textsc{Xillver} for the ionized reflection, and a transmission through a neutral medium. The spectrum below 2~keV was modelled using a simple power law, and a Gaussian component was added to account for the expected neutral iron line at 6.4 keV (fixed). The bottom panel shows the residual between the data and the model in terms of 
$\sigma$. Legend: black and red  refer to the XMM EPIC cameras (pn and the co-added MOS1-MOS2); green refers to \textit{Chandra}; and  light blue, purple, cyan, and yellow refer to the two FPMA and FPMB \textit{NuSTAR} detectors, respectively.}}\label{fig:XMM and Nustar xillver}
\end{figure}

We finally carried out the study of the broad-band properties ($\sim0.5-24$~keV)  of I00183 by modelling  all the spectra together. The three observations were taken at different times (see Table \ref{tab:X-ray observation summary}), but the continuum emission shows similar flux levels: the flux measurements differ  by 20\% at most, but considering that the measurement errors are about 20\%--30\%, we can conclude that there is no significant variability in the continuum flux. We then carried out the joint fitting by grouping all the spectra by 15 counts per bin and using the $\chi^{2}$ statistic. On the basis of the results obtained by modelling the spectra individually, we performed the broad-band analysis using the \textsc{Xillver} model to account for the reflection from an ionized medium, an absorbed power law to model the cold transmission component, a power law modified only by the Galactic absorption to account for the soft excess (below 2 keV), and a Gaussian component to model the expected neutral iron line due to the reflection from cold gas at rest-frame 6.4~keV (fixed). The resulting best-fit spectrum is shown in Figure~\ref{fig:XMM and Nustar xillver} ($\chi^{2}/dof=80/78$). The derived column density of the line-of-sight neutral absorber is $6.8^{+2.1}_{-1.5}\times10^{22}$~cm$^{-2}$, with the photon index fixed at 1.9. The inclination angle was fixed at its hard limit of $87^{\circ}$. The ionization parameter is log$\xi=2.6\pm0.2$. 
%The observed 2$-$10~keV flux is of the order of $(1.44^{+0.35}_{-0.37})\times10^{-13}$~erg~s$^{-1}$~cm$^{-2}$. The de-absorbed 3$-$24~keV luminosity is $(4.2\pm0.1)\times10^{43}$~erg~s$^{-1}$, while that in the 2$-$10 keV range is $\sim(4.2\pm0.1)\times10^{43}$~erg~s$^{-1}$.\\
Table~\ref{tab:X-ray MyTorus parameters} summarizes all the best-fit X-ray parameters derived from \textit{Chandra}, XMM, and \textit{NuSTAR} individual spectral fitting, and from the broad-band analysis. 
%It is worth noting that, the use of self-consistent analytic models, such the recent \textsc{Xillver}, allows the study of the ionized reflection component with greater accuracy with respect to the previous published X-ray study of I00183 \citep[e.g.][]{Nandra07}.

\section{Discussion}\label{sec:discussion}
\subsection{Cold absorber: X-ray versus ALMA}
One of the main aims of our work is to give an estimate of the molecular gas (traced by the CO) column density, which is then compared with that obtained from the X-ray spectral analysis. The purpose is to verify whether the line-of-sight molecular clouds can be considered, at least in part, responsible for the attenuation of the X-ray emission. The galaxy I00183 has been overall classified as an AGN-dominated ULIRG \citep[e.g.][]{Mao14} and selected from the Spoon diagnostic diagram \citep{Spoon07} as a heavily obscured AGN candidate. A CO(1-0) emission line was detected (35$\sigma$) from Band 3 ALMA data, arising from a molecular gas with a total extension (deconvolved from beam) of about $(7.8\pm1.2)\times(5.2\pm2.4)$ kpc$^{2}$. Following the method adopted by \citet{Gilli14}, we assumed a spherical symmetry of the molecular gas regions, taking as radius the median value between the major and the minor axis of the elliptical source size ($r\sim6.4\pm1.7$~kpc). We roughly estimated the column density multiplying the $H_{2}$ density (i.e. $\dfrac{M(H_{2})}{4/3\pi r^{3}}$) by the radius, obtaining N$_{H}=(8.0\pm0.9)\times10^{21}$~cm$^{-2}$. For the X-ray counterpart, the cold gas column density obtained from the best-fit spectral modelling is N$_{H}\simeq6.8^{+2.1}_{-1.5}\times10^{22}$~cm$^{-2}$. The molecular column density derived from the CO is only $\sim$1/8 the column density inferred from the X-ray data; given the uncertainties and the adopted assumptions, this suggests that the line-of-sight molecular gas may contribute   a fraction of the observed AGN obscuration at most. It is also worth noting that several factors can affect our estimate of the molecular gas column density. First, we derived the molecular gas mass using the well-known relation $M(H_{2})=\alpha~L'_{CO}$, with which the CO luminosity (subject to uncertainties related to the observations) is converted into an estimate of the $H_{2}$ mass using the conversion factor $\alpha$. We used the standard conversion factor determined for local ULIRGs \citep[$\alpha=0.8$,][]{Solomon05}, but many uncertainties affect this assumption. The $\alpha$ parameter strongly depends on the galactic environment, varying with metallicity and with density, temperature, and kinetic state of the molecular gas. It ranges from $0.8-1$ for local ULIRGs and starburst galaxies to about 4.8 for `normal' disk galaxies, and its impossible to outline a unique `recipe' for its assumption. For these reasons, the determination of the conversion factor still remains highly uncertain \citep[see][for a review]{Bolatto13}, and its assumption can affect the mass and, in turn, the column density estimation. Second, we measured the line-of-sight molecular gas column density using a very simplified model of a uniform sphere; obviously, this geometrical assumption can affect our column density estimate. In fact, it is highly unlikely that the molecular gas is uniformly distributed within the emitting region; more realistically, the gas distribution varies radially and, if we suppose that our line-of-sight is the one passing through the centre of the sphere and that the gas density decreases outwards, our assumption can lead to underestimating the central column density. %On the other hand, we estimated the line-of-sight column density along the length of the adopted radius; if we suppose that our line-of-sight is the one passing through the center of the sphere, we should take 2 times the radius as the length, adding a factor 2 in our column density estimate. Nevertheless, we considered our assumption a good compromise between the adopted spherical geometry and the elliptical source sizes, taking into account that the length of the line-of-sight passing through the molecular gas distribution could be both longer or shorter than that we assumed.} 
Third, our calculations do not take into account the atomic hydrogen contribution to the gas mass (i.e. M$_{TOT}=$M$_{HI}$+M$_{H2}$) since the amount of M$_{HI}$ is unknown. The lack of this information leads to underestimating the total amount of cold gas and, in turn, the column density. On a purely speculative basis, knowing that M$_{HI}$ is typically 20\%--50\% that of M$_{H2}$ in ULIRGs \citep[e.g.][]{Sanders91}, we can give an estimate of the column density if we include the atomic hydrogen. Reasonably assuming that both atomic and molecular gas are co-spatial, in the most conservative case in which the atomic gas mass is 20\% that of the molecular mass, we can estimate a total gas mass  M$_{HI}$+M$_{H2}$ of about $1.4\times10^{10}$ $M_{\odot}$. Using our simple model, we evaluate a total cold gas column density of about $1.01\times10^{22}$~cm$^{-2}$. Knowing that the molecular gas region extends up to 5~kpc from the nucleus and is probably not uniformly distributed, its contribution to the AGN obscuration, if confirmed by further investigation, could have important implications for the Unified Model, providing new clues about the contribution of the host galaxy to the AGN obscuration.\\
In recent years, many efforts have been made to investigate this issue, both theoretically and observationally. In particular, thanks to the use of the latest generation of telescopes that have revolutionized our knowledge, especially in the millimetre/sub-millimetre  regime, new approaches have been tested and interesting observational evidence has been found about the role of gas and dust at galactic scales in the AGN obscuration \citep[e.g.][Circosta et al., in preparation]{Gilli14, Prieto14, Buchner17a, Buchner17b, Gallerani17}. In particular, \citet{Gilli14} has presented ALMA Cycle 0 continuum observations of a ULIRG at $z=4.75$. Using the same method adopted in our work, from previous [C{\sc iii}] and CO(2-1) lower resolution observations, they derived a galactic scale column density of $\sim1.1\times10^{24}$~cm$^{-2}$ towards the central SMBH. Their finding is that this column density is consistent with that measured from X-ray observations ($N_{H}\sim1.4\times10^{24}$~cm$^{-2}$), concluding that the gas on sub-kiloparsec  scales  may be sufficient to reproduce the observed X-ray obscuration without introducing the presence of a parsec-scale absorber (i.e. the torus).\\
\citet{Prieto14} presented IR Very Large Telescope (VLT) Adaptive Optics (AO) observations of a sample of seven obscured AGN. The high-resolution of the data allows them to resolve the sources on scales down to few tens of parsec at $\lambda=$2$\mu$m, defined by \citet{Prieto10} as the wavelength at which unambiguously identify the point-like source corresponding to the nucleus; in addition, they used two-colour optical HST images to identify also the optical peak (i.e. the optical peak from stellar light) of their galaxy sample and produce reliable dust absorption maps. In all galaxies they found the presence of dust from the central parsec to the kiloparsec scales; its position in front of the nucleus leads them to consider it being the major cause of the AGN obscuration, justifying the type-II classification. They also took into account that a torus-like inner obscuring structure can form from the flow of material from the outer part of the galaxy to the centre, but they concluded that, for what concerns hiding the nucleus, the larger scale line-of-sight dust structures may be sufficient.\\
Recently, \citet{Buchner17a} presented an X-ray study of a large sample of \textit{Swift}-detected long-duration gamma-ray bursts and investigated their line-of-sight column densities. From the geometry of the obscurer and comparison with local galaxies, they argued that the obscuring medium is consistent with gas on galactic scales; they also find  a power law relation between the obscuring column density and the galaxy stellar mass (i.e. $N_{H} \propto M_{\bigstar}^{1/3}$). In a following paper, \citet{Buchner17b} applied their model to a simulated AGN population, finding that galaxy-scale obscuration is sufficient to explain the obscured AGN fractions at $N_{H}=10^{22}-10^{23}$~cm$^{-2}$, while it fails to provide heavier obscuration ($N_{H}>10^{23}$~cm$^{-2}$) for which a nuclear obscurer is required. They also argued that their results were consistent with that obtained by \citet{Maiolino95} which stated that type-II AGN are preferentially found in edge-on galaxies. Moreover, \citet{Buchner17b} investigated the incidence of the nuclear obscurer by subtracting galaxy-scale obscuration, and found a persistent 40\% galaxy-scale obscurer (at Compton-thin column densities), while the nuclear obscurer is luminosity-dependent and disappears at $L_{2-10~keV}>10^{44.5}$~erg~s$^{-1}$.\\
\citet{Gallerani17} presented \textit{Chandra} observations of a z=6.4 quasar. By using the model presented in \citet{Gallerani14}, they computed the intrinsic X-ray flux emitted by the quasar assuming that the X-ray emission is attenuated by photoelectric absorption due to line-of-sight molecular clouds distributed on kiloparsec scales. They estimated a molecular column density of the order of $10^{22}$~cm$^{-2}$, finding that the estimated amount of molecular gas mass (from which the column density was derived) was in good agreement with that obtained from previous studies.\\
From the cases discussed above, it becomes clear that the galaxy-scale gas can have an important role in AGN obscuration, at least at intermediate (\textit{Compton-thin}) column densities. This provides useful support to our hypothesis of a contribution of kiloparsec-scale molecular gas to the obscuration. Nevertheless, it is worth noting that appropriate considerations about the geometry of the obscuring medium are fundamental for these kind of studies, but in most cases they are not fully supported by the data quality; this factor raises the uncertainties in this type of analysis. In particular, the limited statistics of the X-ray data and the low spectral and spatial resolution of the ALMA data used in this work introduce numerous uncertainties in the estimated quantities, preventing us from drawing strong conclusions. Therefore, our interpretation remains purely speculative and needs to be tested with data of better quality; at the same time, our work highlights that a multi-wavelength approach, using the best data currently available, is likely to be the key strategy used to investigate these issues.\\
Finally, we note that although our column density estimate is comparable with that of a typical Milky Way giant molecular cloud \citep[$\sim10^{22}$~cm$^{-2}$, e.g.][]{Heyer15}, it might be a chance coincidence. Our limited spatial resolution means that the higher column density (N$_{H} \sim 10^{23}$~cm$^{-2}$) inferred from X-ray \citep[and silicate absorption;][this work]{Spoon04} is for a compact obscuring region near the central source and the surrounding region of lower density molecular gas might dilute our measurement. %our column density estimate is comparable with that of a typical Milky Way giant molecular cloud \citep[$\sim10^{22}$~cm$^{-2}$, e.g.][]{Heyer15}, suggesting that an alignment by chance between a cloud and the AGN could provide the necessary obscuration. However, our limited spatial resolution and the distance of our object rule out the possibility to investigate the geometry and distribution of the molecular gas, preventing us from exploring this hypothesis.

\subsection{AGN-like excitation?}
In spite of the limited S/N of the HCN(4-3) and the upper limit given for the HCO$^{+}$(4-3) and HNC(4-3) detections , it is worth briefly discussing them.\\
Hydrogen cyanide (HCN) is one of the most abundant molecules that traces \textit{dense} molecular gas ($n>3\times10^{4}$~cm$^{-3}$) directly associated with active star-forming regions \citep[e.g.][]{Carilli13}, and is often used along with other gas tracers to probe the excitation conditions of the molecular gas. In particular, HCN, HCO$^{+}$, and CO line ratios have been used in several works as powerful tools to investigate  the relative contribution of star formation and AGN to the excitation of the interstellar medium  \citep[e.g.][and references therein]{Imanishi06,Aalto07,Krips10,Izumi13,Imanishi14,Izumi16,Privon17}. Emission lines at millimetre wavelengths are only marginally affected by dust extinction, making them the preferential method used to assess the dominant energy source in the inner regions of dust-rich galaxies such as ULIRGs. The idea is that the radiation field produced by the AGN impacts the surrounding medium, developing the so-called X-ray dissociation regions \citep[XDRs][]{Maloney96}; the analogous expected in starburst regions are the photodissociation regions (PDRs). Since X-rays from AGN penetrate deep into the surrounding medium, the resulting XDRs are expected to be large and to affect the molecular gas properties on large, host-galaxy scales \citep[e.g.][]{Aalto07}. As a consequence, different flux ratios are expected in XDRs and PDRs. In particular, HCO$^{+}$ molecules are though to be destroyed in XDRs, while HCN molecules are believed to maintain high abundances, then enhanced HCN/HCO$^{+}$ and HCN/HNC flux ratios are expected in AGN rather than starburst-heated molecular gas; as a consequence, enhanced HCN emission has been used to detect AGN in ULIRGs \citep[e.g.][]{Imanishi06,Imanishi14}.\\ %However, several authors stated that one needs to be cautious on the interpretation of these line ratios. One one hand, because some of them have found enhanced HCN emissions in pure starburst (or at most composite) rather than AGN  systems \citep[e.g.][]{Krips08,Costagliola11}; on the other hand, because the processes driving enhanced HCN abundances may be different \citep[i.e. mechanical heating by supernovae or feedback from AGN, e.g.][]{Privon17} and their interpretation is not always straightforward. Nevertheless, investigating the molecular line ratios can provide useful constraints on the excitation conditions of the molecular gas in the inner part of a galaxy.\\
We detected a 6$\sigma$ HCN emission with an integrated flux density of $\sim$0.4~Jy~km~s$^{-1}$ (in a $\sim$66~km~s$^{-1}$ line FWHM). Assuming that HCO$^{+}$ and HNC emissions may have the same FWHM as the HCN, we derived 3$\sigma$ upper limits for the HCO$^{+}$ and HNC integrated intensities of $<0.2$~Jy~km~s$^{-1}$ and $<0.1$~Jy~km~s$^{-1}$, respectively. This results in HCN/HCO$^{+}>2$ and HCN/HNC$>3$. The obtained lower limits for the line ratios are in the range usually found for AGN-like excitations \citep[e.g.][]{Imanishi14,Izumi16}. At the observed kiloparsec scales, AGN-like line ratios would suggest that the radiation is escaping the vicinity of the AGN and is interacting with the molecular gas at larger scales. This is likely to provide supporting evidence to a possible contribution of the galactic-scale gas to the observed AGN obscuration, but given the uncertainties affecting our estimates, we cannot draw solid conclusions.

\section{Summary and conclusions}\label{sec:concl}
In this paper we have presented a study of the multi-frequency properties of the ULIRG IRAS 00183-7111 at z$=0.327$; one of the main goals was to estimate at what level the molecular gas traced by the CO can be responsible for the obscuration observed in X-ray. To link the millimetre to the X-ray properties of I00183, ALMA archival Cycle 0 observations in Band 3 ($\sim$87~GHz) and Band 6 ($\sim270$~GHz) were calibrated and analysed. The main results obtained analysing ALMA data can be summarized as follows:
\begin{itemize}
\item From the 35$\sigma$ CO(J=1-0) detection, we measured a molecular gas mass M$_{H_{2}}=(1.14\pm0.10)\times10^{10}$~M$_{\odot}$, obtained assuming a H$_{2}$ mass-to-CO luminosity conversion factor typical for ULIRGs (i.e. $\alpha=0.8$).
\item The correlation between the CO and the FIR luminosity was used to compute a star formation rate of $\sim260\pm28$~M$_{\odot}$~yr$^{-1}$, comparable (within the errors) with that obtained from an ATCA CO detection by \citet{Mao14}, using the same method.
\item A blueshifted excess was found in the CO integrated spectral profile that was interpreted as a hint of the presence of an outflow from the CO emitting region. The CO channel map also shows a eastward protrusion extending up to 7'' from the nucleus which seems to be spatially coincident with the brighter knot of the [O{\sc iii}] outflow discussed by \citet{Iwasawa17}. However, the poor spectral (90~km~s$^{-1}$) and spatial ($\sim3"$) resolutions of the data do not allow us to draw strong conclusions. 
\item By adopting a simple spherical model for the molecular cloud, the molecular gas mass was used to derive an estimate of molecular gas column density along the line of sight: $N(H_{2})\sim(8.0\pm0.9)\times10^{21}$~cm$^{-2}$.
\item We obtained three continuum maps at 87, 260, and 270~GHz. The corresponding integrated flux densities shows a flat spectral index $\alpha\simeq-0.24$ ($S_{\nu} \propto \nu^{\alpha}$). We fitted the FIR spectral energy distribution of I00183 (based on Herschel flux densities) with a modified black body to estimate the contribution of thermal dust emission at the observed ALMA frequencies. $T_{d}=68\pm10$~K and $M_{d}=(2.7\pm1.4)\times10^{7}$~M$_{\odot}$ were obtained as best-fit parameters. This allows us to estimate optically thin synchrotron emission and thermal emission from dust as the dominant emission mechanism at 87~GHz and 270~GHz, respectively. The flat spectral index between these two frequencies likely arises from a mixture of processes, including free-free emission from star-forming regions.
\item  An offset of $\sim2.5$~kpc was found between the CO line and the continuum emission at a $7\sigma$ significance. The interpretation of this shift is not straightforward, mainly because of the limited data resolution. Nonetheless, we speculate about a possible origin due to an interaction between the molecular gas region and the  radio-loud nuclear source recently observed at radio wavelengths by \citet{Norris13}. 
\item We detected a 6$\sigma$ HCN(J=4-3) emission and estimated 3$\sigma$ upper limits of HCO$^{+}$(J=4-3) and HNC(J=4-3) emissions. They were used to derive lower limits of the well-known molecular line ratios, used to assess the relative contribution of star formation and AGN to the ISM excitation. Our estimates suggest an AGN-like excitation.
\end{itemize}
The X-ray analysis was carried out using \textit{Chandra}, XMM-\textit{Newton}, and newly acquired \textit{NuSTAR} data, allowing  broad-band coverage of the X-ray spectrum (0.5$-$24~keV, observed frame). The spectra were first analysed individually, showing similar spectral shape and consistent observed fluxes (within the errors). All the spectra were then used for a simultaneous X-ray spectral fitting. The main X-ray results can be summarized as follows:
\begin{itemize}
\item The detection of the He-like iron line (Fe XXV) suggests that the reflection arises from a strongly ionized medium. Using the \textsc{Xillver} model, we derived an ionization parameter log$\xi=2.6$, which is consistent with the presence of a powerful source of radiation that strongly ionizes the surrounding medium. 
\item The photon index is consistent with that expected from theory ($\sim1.8/1.9$), with an intrinsic N$_{H}$ column density of $\sim6.8\times 10^{22}~cm^{-2}$.
\end{itemize}
The column densities derived separately in the millimetre and X-ray bands were then compared to assess the possible contribution of the molecular gas to the AGN obscuration. The molecular gas column density is $\sim$1/8 of that obtained from X-ray analysis, but given the uncertainties and the adopted assumptions, suggests that the gas on galactic scales may contribute to a fraction of the AGN obscuration; however,  the link is not straightforward. The inner regions of I00183 are likely stratified into different layers of matter, extending from a powerful nuclear engine (intrinsic $L_{2-10 keV}>10^{44}$ erg s$^{-1}$); the inner regions are highly ionized by its strong radiation field, while the surrounding layers are composed of colder gas, which extends up to 5~kpc from the nucleus and is responsible of a vigorous star formation rate of $260$~M$_{\odot}$~yr$^{-1}$. The complexity of this nuclear environment makes it hard to assess the origin of the AGN obscuration with the spatial and spectral resolution of the available data. We may suppose that in such an environment the obscuration can also occur  at galactic scales, but currently most of our claims remain purely speculative and need to be tested with higher resolution observations. Nevertheless, our work shows how studies like these potentially provide useful constraints on the conditions of the central regions of a source, allowing us to investigate  the properties of the AGN obscuration and to better understand the nature of local ULIRGs.
%We can conclude that the interplay between the AGN and its host galaxy seems to be crucial to explain lots of the observed AGN properties. Similar results highlight the importance of these studies, not only to better understand the nature of local ULIRGs and investigate about the properties of the AGN obscuration, but also to improve our knowledge about the interplay and mutual feedback between star formation and black hole accretion, which are basic ingredients of galaxy formation and evolution. 

\begin{acknowledgements}
    We thank the anonymous referee for the helpful comments that lead to significantly improving this paper. We thank Dr. T. A. Davis for making available his routines for producing quality moment maps and plots and providing useful discussions about the argument. We also thank Dr. Andrea Giannetti for valuable discussions about dust.\\
        This paper makes use of the following ALMA data: ADS/JAO.ALMA\#[2011.0.00034.S] and  ADS/JAO.ALMA\#[2011.0.00020.S]. ALMA is a partnership of ESO (representing its member states), NSF (USA), and NINS (Japan), together with NRC (Canada), NSC and ASIAA (Taiwan), and KASI (Republic of Korea), in cooperation with the Republic of Chile. The Joint ALMA Observatory is operated by ESO, AUI/NRAO, and NAOJ. The scientific results reported in this article are also based on observations made by Chandra X-ray Observatory, NuSTAR, and XMM-Newton, and has made use of the NASA/IPAC Extragalactic Database (NED) which is operated by the Jet Propulsion Laboratory, California Institute of Technology under contract with NASA. This research has also made use of data and software provided by the High Energy Astrophysics Science Archive Research Center (HEASARC), which is a service of the Astrophysics Science Division at NASA/GSFC and the High Energy Astrophysics Division of the Smithsonian Astrophysical Observatory. 
\end{acknowledgements}

\bibliographystyle{aa} % style aa.bst
\bibliography{mybib} % your references Yourfile.bib

\end{document}